\documentclass[%
 reprint,
 amsmath,amssymb,
 aps,
]{revtex4-2}

\usepackage{natbib}
\usepackage{graphicx}
\usepackage{dcolumn}
\usepackage{bm}

\begin{document}

\preprint{APS/123-QED}

\title{Generation and entanglement study of generalized $N$-mode single photon perfect W-states}

\author{Manoranjan Swain}
\email{swainmanoranjan333@gmail.com}
\affiliation{Department of Physics and Astronomy, National Institute of Technology, Rourkela, 769008, Odisha, India}

\author{M. Karthick Selvan}
\email{karthick.selvan@yahoo.com}
\affiliation{Department of Physics, Thiagarajar College, Madurai-625009, Tamilnadu, India}

\author{Amit Rai}
\email{amitrai007@gmail.com}
\affiliation{School of Physical Sciences, Jawaharlal Nehru University, New Delhi 110067, India}

\author{Prasanta K. Panigrahi}
\email{pprasanta@iiserkol.ac.in}
\affiliation{Department of Physical Sciences,\\ Indian Institute of Science Education and Research Kolkata, Mohanpur 741246, West Bengal, India}


\begin{abstract}
We consider single photon realization of generalized $N$-qubit perfect W-states which are suitable for perfect teleportation and superdense coding. We propose schemes to generate generalized $N$-mode single photon perfect W-states and derive entanglement conditions which for single photon states require finding fidelity with generalized $N$-mode single photon perfect W-states and hence more suitable to detect the genuine entanglement of generalized perfect W-states. Based on the evolution of single photon wavefunction in scalable integrated photonic lattices, we present schemes for the preparation of generalized $N$-mode single photon perfect W-states at desired propagation distance. The integrated waveguide structures can precisely be fabricated, offer low photon propagation losses and can be integrated on a chip. We consider both planar and ring type waveguide structures for state generation. We derive set of generalized entanglement conditions using the sum uncertainty relations of generalized $su(2)$ algebra operators. We show that any given genuinely entangled $N$-mode single photon state is a squeezed state of a specific $su(2)$ algebra operator and can be expressed as superposition of a pair of orthonormal generalized $N$-mode single photon perfect W-states which are eigenstates of that specific $su(2)$ algebra operator. Within the single photon subspace, the eigendecomposition of $su(2)$ algebra operators reduces the generalized entanglement condition to a simplified single photon separability condition. In order to verify the entanglement of given genuinely entangled $N$-mode single photon state using this condition one has to find the difference between the state fidelities with suitably chosen pair of orthonormal generalized $N$-mode single photon perfect W-states. Finally, we propose an experimental scheme to verify the entanglement using the proposed conditions. This scheme uses a photonic circuit that consists of directional couplers, and phase shifters and the same photonic circuit can also be used to generate generalized $N$-mode single photon perfect W-states. Our results show that optical waveguide structures
are ideal platforms for generation and entanglement verification of large $N$-mode single photon entangled states which could find novel applications in quantum
information science. 
\end{abstract}

\maketitle


\section{Introduction}

Multipartite entangled states~\cite{borras2007multiqubit,facchi2008maximally,enriquez2016maximally} involve complex entanglement structure~\cite{miyake2003classification,koashi2000entangled} and have drawn wide interest recently. In the case of three qubits, there exist two in-equivalent classes of maximally entangled states, known as the GHZ-class~\cite{kafatos2013bell} and the W-class states~\cite{dur2000three}. For more than three qubits, we have other interesting entangled states~\cite{dur2003multiparticle,hein2004multiparty,nielsen2006cluster}. These multipartite entangled states are useful for various applications in quantum information processing~\cite{kempe1999multiparticle,yeo2006teleportation}. One of such applications is quantum teleportation~\cite{pirandola2015advances,muralidharan2008perfect,choudhury2009quantum,saha2012n}, where entangled state is shared as quantum channel between the sending and receiving ends. Teleportation with Bell states~\cite{bennett1993teleporting}, GHZ states~\cite{karlsson1998quantum} and other multipartite states~\cite{kumar2020experimental} were shown to be possible. Among the multipartite entangled states, W-class states are known for their robustness against particle loss. A set of states belonging to W-class can be used for perfect teleportation~\cite{agrawal2006perfect,rao2008generation}. The generalization of such states for $N$-qubit system is given as (up to global phase),

\begin{equation*}
    \vert \Tilde{W} \rangle_N = \dfrac{1}{\sqrt{2}} \bigg[ \sum_{j=1}^{N-1} \alpha_j  \vert 1 \rangle_{j} \vert 0...0\rangle_{1..j-1,j+1..N} \bigg]
\end{equation*}
\begin{equation}\label{eq1}
\begin{aligned}
 + \dfrac{1}{\sqrt{2}} \vert 00....01 \rangle_{1...N}
\end{aligned}
\end{equation}

where $\alpha_j$'s are non-zero complex coefficients such that $\vert \alpha_j \vert < 1$ and $\sum_{j=1}^{N-1} \vert \alpha_j \vert^2 = 1$. 

We denote such states as generalized $N$-qubit perfect W-states. When $\alpha_j= \dfrac{1}{\sqrt{N-1}}$, $\forall j$, the state $\vert \Tilde{W} \rangle_N$ reduces to 

\begin{equation*}
\vert W\rangle_N=\dfrac{1}{\sqrt{2(N-1)}}\bigg[ \sum_{j=1}^{N-1} \vert 1 \rangle_{j} \vert 0...0\rangle_{1..j-1,j+1..N} \bigg]
\end{equation*}
\begin{equation}\label{eq2}
\begin{aligned}
+\dfrac{1}{\sqrt{2}}\vert 00....01\rangle_{1...N},
\end{aligned}
\end{equation}

 which has been shown to be useful for perfect teleportation and superdense coding~\cite{rao2008generation,li2007states,li2016generating}. We refer this state as $N$-qubit perfect W-state. Although the state holds important applications in quantum information processing, generation of such states with more number of qubits has been a challenging task. A scheme using exchange interaction between electron spins was proposed to generate this state in quantum dots~\cite{rao2008generation}. However, in this system, it is difficult to preserve multiqubit entanglement due to the decoherence of electron spins. A cavity QED scheme which uses two energy levels of atoms as qubits was proposed to prepare perfect W-state~\cite{zhao2008scheme} and superconducting qubits were used to generate 3-mode perfect W-states~\cite{swain2020generation}. However, these systems are not scalable to generate large $N$-qubit perfect W-states. In another scheme, fusion and expansion mechanisms were used to prepare large $N$-qubit perfect W-states~\cite{li2016generating}. This scheme uses polarization degree of freedom of photons as qubits and requires small sized perfect W-states as initial states.  
 
  A promising system to realize generalized $N$-qubit perfect W-states is a single photon shared between $N$ spatial modes~\cite{van2005single,morin2013witnessing,shi2013heralded,grafe2014chip,monteiro2015revealing,caspar2020heralded,perez2013generating,hessmo2004experimental}. Single photon path entangled states have been reported to be useful in quantum random number generation~\cite{white2020quantum,chen2019single,luo2020quantum}, quantum repeater~\cite{gottesman2012longer},  optical Bloch oscillation~\cite{rai2009possibility}, to name a few examples. Optical photonic waveguides can readily be used to generate such single photon entangled states. Integrated waveguide lattices are gaining interest for their diverse applications in physics. These are scalable, offer low loss and can be precisely fabricated by femtosecond laser direct writing technique~\cite{szameit2010discrete,meany2015laser}. They are compact and are realizable with current technologies, and hence can serve as an important tool for optical simulation~\cite{keil2015optical,rai2015photonic} and generation of entangled states~\cite{grafe2014chip,perez2013generating,heilmann2015novel,rai2010quantum}. 

In this paper, we propose schemes to generate generalized $N$-mode single photon perfect W-states using weakly coupled waveguide structures. In our previous work, we proposed scheme to generate generalized 3-mode single photon perfect W-states using 1-dimensional (1D) integrated waveguide structures~\cite{swain2020single}. Here we consider both 1D structure (planar structure) and 2-dimensional (2D) ring structure of single mode waveguides. In the 1D structure, the spacing between two successive waveguides is kept different to ensure different coupling strengths between waveguides. In 2D ring structure, $N$ waveguides are symmetrically arranged to form a ring and another waveguide is kept at the center of ring. The same geometries were considered to propose schemes for single photon symmetric W-state generation~\cite{perez2013generating}. However, the generation of generalized $N$-mode single photon perfect W-states requires completely different coupling schemes and hence different waveguide structures. Specifically, in our scheme to generate generalized $N$-mode single photon perfect W-states using 2D ring structure, it is possible to have more than 6 waveguides on the ring. However, in the case of generation of single photon symmetric W-state, one can have only 6 waveguides on the ring~\cite{perez2013generating}.  

Verification and quantification of entanglement of generated states are very crucial for many quantum information processing applications. Two-qubit entanglement can be quantified using entanglement measures~\cite{wootters1998entanglement,horodecki2009quantum}. Extensions of bipartite entanglement measures~\cite{ma2011measure,rafsanjani2012genuinely} and monogamy relations~\cite{coffman2000distributed,koashi2004monogamy,zhu2015generalized} are used to study the multipartite entanglement. Entanglement detection conditions can be used to distinguish entangled states from separable states. For two-qubit systems, the positivity of partial transposition (PPT) is a both necessary and sufficient condition for separability~\cite{peres1996separability,horodecki1996necessary}. For multiqubit systems, entanglement detection conditions based on entanglement witness operators~\cite{terhal2002detecting,bourennane2004experimental,guhne2009entanglement} and spin squeezing inequalities~\cite{guhne2009entanglement} can be used to distinguish genuinely entangled states from separable states. Especially the genuine entanglement of single photon states can be detected using entanglement detection conditions~\cite{agarwal2005inseparability,hillery2006entanglement,hillery2006entanglement1,nha2006entanglement} which involves violation of sum and product uncertainty relations constructed using $su(2)$ and $su(1,1)$ algebra operators. These conditions can be experimentally verified~\cite{nha2006entanglement}. Entanglement witness operators can also be used to detect the entanglement of multimode single photon states~\cite{monteiro2015revealing,caspar2020heralded,nha2008linear}. 

Single photon perfect W-states, generated by injecting single photon through coupled waveguide structures, are not biseparable and retain bipartite entanglement between each pairs of modes. We derive a set of generalized entanglement detection conditions based on sum uncertainty relation of generalized $su(2)$ algebra operators to detect the entanglement of generalized $N$-mode single photon perfect W-states. Recently proposed entanglement condition~\cite{selvan2019entanglement} to detect the entanglement of multimode W-type entangled states belongs to this set of conditions. In the case of single photon states, we show that the genuinely entangled $N$-mode single photon states are squeezed states of $su(2)$ algebra operators and reduce the generalized entanglement detection condition to a single photon separability condition which involves only the fidelities of the generated $N$-mode single photon state with two pairs of orthonormal generalized $N$-mode single photon perfect W-states. These generalized $N$-mode single photon perfect W-states are eigenstates of $su(2)$ algebra operators involved in the generalized entanglement detection condition. We propose an integrated photonic circuit~\cite{matthews2009manipulation} consisting of directional couplers and phase shifters for experimental implementation of generalized entanglement detection conditions. It involves detecting the sum and difference of photon numbers at two specific outputs of the photonic circuit. However, entanglement verification of genuinely entangled single photon states requires detecting only the photon number difference at the two specific outputs. Finally, we show that the same photonic circuit can also be used to generate generalized $N$-mode single photon perfect W-states. Thus the waveguide structures are promising systems to generate and verify the entanglement of large $N$-mode generalized single photon perfect W-states.  

 Our paper is organized as follows. In section 2, we describe the generation of generalized $N$-mode single photon perfect W-states using 1D-planar waveguide structure. In section 3, the generation of that state using 2D-ring structure is explicated. In section 4, we derive the generalized entanglement conditions using sum uncertainty relation of $su(2)$ algebra operators and reduce it to single photon separability condition. In section 5, the experimental implementation of proposed entanglement conditions using an integrated photonic circuit is explained. Finally, we arrive at conclusion in section 6. 
 
\section{Generation of generalized $N$-mode single photon perfect W-states using 1D structure}
\begin{figure}[h]
\centering
\includegraphics[scale=0.7]{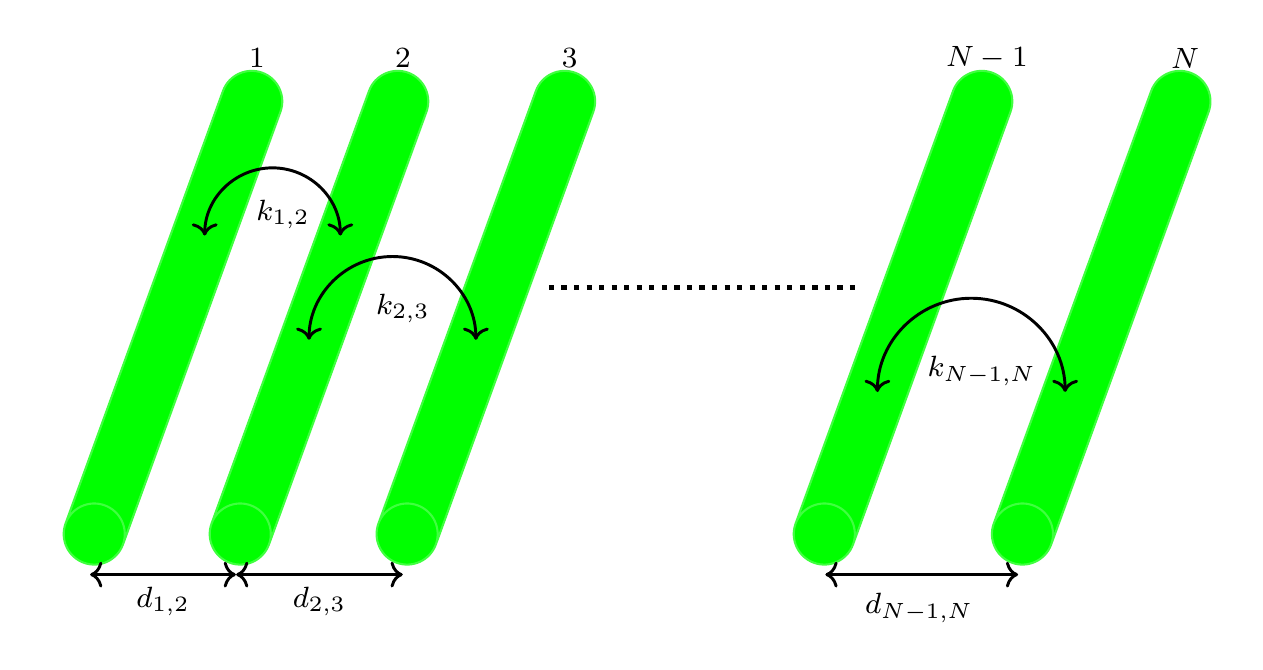}
\caption{One dimensional array of $N$ identical waveguides. $k_{i,j}$ and $d_{i,j}$ are coupling strength and separation between waveguides $i$ and $j$ respectively.}
\label{Fig. 1}
\end{figure}

 For one dimensional array of $N$ identical waveguides, as shown in Figure \ref{Fig. 1}, the Hamiltonian can be written as,
\begin{equation}
 \hat{H}=\hbar\omega\sum_{j=1}^N\hat{a}_j^\dagger\hat{a}_j+ \hbar \sum_{j=1}^{N-1}k_{j,j+1}(\hat{a}_j^\dagger \hat{a}_{j+1}+\hat{a}_j \hat{a}_{j+1}^\dagger )
\end{equation}
where the first term represents the free propagation of light with $\omega$ proportional to refractive index of the material. $\hat{a}_j^\dagger(\hat{a}_j)$ is the bosonic creation(annihilation) operator. $k_{j,j+1}$ is the coupling strength between waveguides $j$ and $j+1$ which depends on the separation ($d_{j,j+1}$) between them. The Heisenberg equations of motion are given by,

\begin{equation*}
    i \frac{d\hat{a}_1^{\dagger}}{dz} = k_{1,2} \hat{a}_2^{\dagger} ,
\end{equation*}

\begin{equation*}
 i \frac{d\hat{a}_j^\dagger}{dz}= k_{j-1,j} \hat{a}_{j-1}^\dagger+ k_{j,j+1}\hat{a}_{ j+1}^\dagger,~~(j=2,...N-1)
\end{equation*}

\begin{equation}\label{m}
 i \frac{d \hat{a}_N^{\dagger}}{dz} = k_{N-1,N} \hat{a}_{N-1}^{\dagger}.
\end{equation}
The solution to the above set of equations can take the form,
\begin{equation}\label{eq4}
 \hat{A}^\dagger(z)=e^{-izM}\hat{A}^\dagger(0)
\end{equation}

where $\hat{A}^{\dagger}$ is the column of creation operators, $M$ is the coupling matrix and $e^{-izM}$ is the evolution matrix. For given number of waveguides with appropriate coupling strengths between them, Eq. \ref{eq4} can be used to find the value of $z$ for which the initial state evolves to generalized $N$-mode single photon perfect W-state. In the following we consider the generation of $4$-mode and $5$-mode single photon perfect W-states. 

The 4-mode single photon perfect W-state is given as,
\begin{equation}\label{eq3}
  \vert W\rangle_4= \frac{1}{\sqrt{6}
  }(\arrowvert1000\rangle+ \arrowvert0100\rangle+ \arrowvert0010 \rangle + \sqrt{3}\arrowvert0001\rangle)
\end{equation}

To generate 4-mode single photon perfect W-state, we consider an array of 4 waveguides in which a single photon is injected in the third waveguide. The equations of motion can be written in the matrix form as follows.

\begin{equation}{\label{eq6}}
i\frac{d}{dz}
 \begin{pmatrix}
\vspace{0.2cm} a_1^\dagger \\ \vspace{0.2cm} a_2^\dagger\\a_3^\dagger\vspace{0.2cm}\\a_4^\dagger

 \end{pmatrix}
= 
\begin{pmatrix}
 0 & k_{1,2} & 0&0\\k_{1,2} & 0 & k_{2,3} &0\\0 & k_{2,3} & 0 & k_{3,4}\\ 0&0&k_{3,4}&0
\end{pmatrix}
\begin{pmatrix}
 \vspace{0.2cm} a_1^\dagger \\ \vspace{0.2cm} a_2^\dagger\\a_3^\dagger\vspace{0.2cm}\\a_4^\dagger

\end{pmatrix}
\end{equation}

The initial state of the system is $\vert \psi(0) \rangle = \hat{a}_3^\dagger \vert 0000 \rangle$. After propagating a distance `$z$' the state evolves to $|\psi(z)\rangle$= $C_{1}|1000\rangle$ + $C_{2}|0100\rangle$ + $C_{3}|0010\rangle$ +
$C_{4}|0001\rangle$. To obtain the desired state, the parameters $z$ and $k_{i,j}$ are to be selected in such a way that $|C_1|^2=\frac{1}{6}$, $|C_2|^2=\frac{1}{6}$,  $|C_3|^2=\frac{1}{6}$ and $|C_4|^2=\frac{1}{2}$. For the values of coupling constants and distance $z$ listed in TABLE \ref{parmeters}, the following generalized 4-mode single photon perfect W-state can be obtained by solving the above matrix equation. 

\begin{equation}\label{eq13}
 \arrowvert \Tilde{W}\rangle_4= \frac{1}{\sqrt{6}}(-\arrowvert 1000\rangle +i\arrowvert0100\rangle + \arrowvert 0010\rangle+ i \sqrt{3}\arrowvert 0001\rangle)
\end{equation}

\begin{table}[h]
 \centering
  \caption{The parameters required for generation of 4-mode and 5-mode single photon perfect W-states }
 \begin{tabular}{c|c|c|c|c|c}
 \hline \hline
 No. of &  $k_{1,2}$ &  $k_{2,3}$&  $k_{3,4}$ &  $k_{4,5}$  &  $z$ \\
 modes & $(cm^{-1})$ & $(cm^{-1})$ & $(cm^{-1})$ & $(cm^{-1})$ & $(cm)$ \\
 \hline
 & & & & & \\
4   & 1.2043   & 0.686372  & 0.781121  & - & 1.15042 \\ & & & & & \\
 \hline
 & & & & & \\
5  & 1.08983   & 0.584456  & 0.988893  & 1.53062 & 1.23828\\ & & & & & \\
         
 \hline \hline
 \end{tabular}
 \label{parmeters}
\end{table}

 The 4-mode single photon perfect W-state can be obtained by adjusting the phase terms of individual ket vectors in Eq.$\eqref{eq13}$. This phase adjustment can be performed by adding phase shifters~\cite{matthews2009manipulation} to individual guides. In this particular case, phase shifters have to be added to provide a phase shift of $`-1$' in the $1^{\text{st}}$ waveguide and of $`-i$' in the $2^{\text{nd}}$ and $4^{\text{th}}$ waveguides. 
 
 Next we consider 5-mode single photon perfect W-state which can be written as,

\begin{equation}\label{eq3a}
  \vert W\rangle_5= \frac{1}{2\sqrt{2}
  }(\arrowvert10000\rangle+ \arrowvert01000\rangle+ \arrowvert 00100\rangle + \arrowvert 00010\rangle+ 2\arrowvert00001\rangle)
\end{equation}

This state can be generated using an array of 5 waveguides. The equations of motion can be written and solved as described above. Assuming that the photon is injected in the central waveguide, for the values of coupling parameters and distance $z$ given in TABLE \ref{parmeters}, the following generalized 5-mode single photon perfect state can be obtained.

\begin{equation*}
\vert \Tilde{W}\rangle_5= \frac{1}{2\sqrt{2}
  }(-\arrowvert10000\rangle+ i \arrowvert01000\rangle+ \arrowvert 00100\rangle
\end{equation*}
\begin{equation}\label{eq3b}
\begin{aligned}
   + i \arrowvert 00010\rangle-2\arrowvert00001\rangle)
\end{aligned}
\end{equation}

In order to get the 5-mode single photon perfect W-state, the phase shifters have to be added to provide phase shift of `$-1$' in the $1^{\text{st}}$ and $5^{\text{th}}$ waveguides and of `$-i$' in the $2^{\text{nd}}$ and $4^{\text{th}}$ waveguides. 

In similar way, it is possible to have coupling scheme for any $N$ number of waveguides arranged in planar structure to generate generalized $N$-mode single photon perfect W-states.

\section{Generation of generalized $N$-mode single photon perfect W-states using 2D structure}
\begin{figure}[h]
\centering
\includegraphics[scale=0.7]{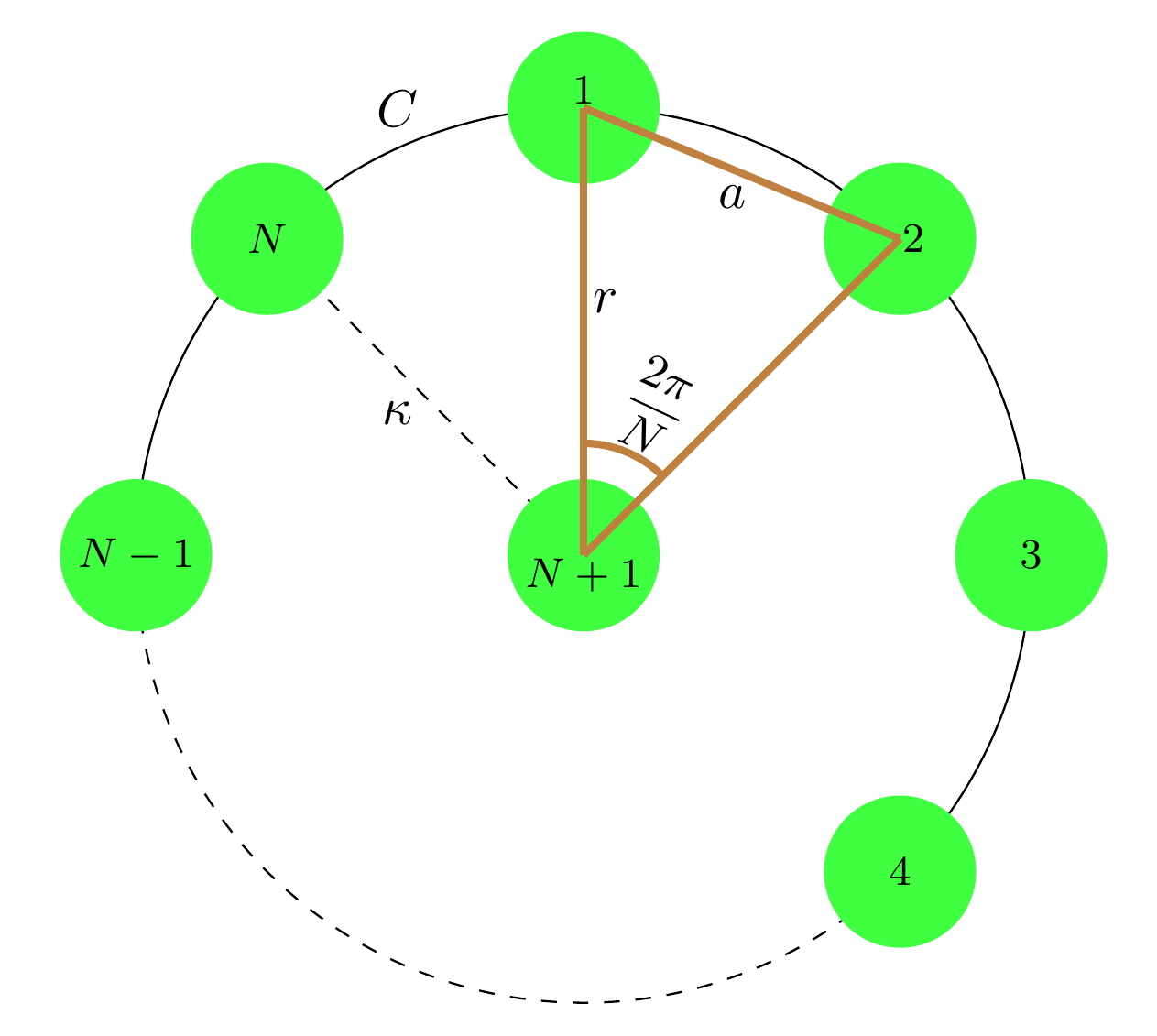}
\caption{$N+1$ waveguides arranged in 2D ring structure. The surrounding waveguides are labelled by integers from $1$ to $N$ and the central waveguide is labelled as $N+1$. $r$ and $a$ are radius of the ring and the nearest neighbor distance between surrounding waveguides respectively. The angular separation between two nearest surrounding waveguides is $2\pi/N$}
\label{Fig.2}
\end{figure}

In the following, we consider $N+1$ identical waveguides, one at the center and $N$ waveguides arranged symmetrically around it, as shown in Figure \ref{Fig.2}. The central waveguide is coupled with all surrounding waveguides with coupling strength $\kappa$ and each surrounding waveguide is coupled with nearest neighbors with coupling strength $C$. This type of two dimensional ring structure can be fabricated precisely using femtosecond laser direct writing technique~\cite{szameit2010discrete,meany2015laser}. A photon is injected at the central waveguide. 

The Hamiltonian of this system is 

\begin{equation*}
\hat{H}=\hbar\omega \hat{a}_{N+1}^\dagger\hat{a}_{N+1}+\hbar\omega\sum_{j=1}^{N}\hat{a}_j^\dagger\hat{a}_j+ \hbar \kappa \sum_{j=1}^{N}(\hat{a}_{N+1}^\dagger \hat{a}_{j}+\hat{a}_{N+1} \hat{a}_{j}^\dagger )  
\end{equation*}
\begin{equation}\label{eq10}
+ \hbar C \bigg[\hat{a}^{\dagger}_1 \hat{a}_N + \hat{a}_1 \hat{a}^{\dagger}_N + \sum_{j=2}^{N}(\hat{a}_j \hat{a}^{\dagger}_{j-1} + \hat{a}^{\dagger}_j \hat{a}_{j-1}) \bigg],
\end{equation}

with Heisenberg's equations of motion 
\begin{equation*}
    i \frac{d \hat{a}_{N+1}^\dagger}{dz}= \kappa \sum_{j=1}^{N} \hat{a}_j^\dagger,
\end{equation*}
\begin{equation*}
    i \frac{d \hat{a}_1^\dagger}{dz}= C (\hat{a}^{\dagger}_2 + \hat{a}^{\dagger}_N) + \kappa \hat{a}^{\dagger}_{N+1}, 
\end{equation*}
\begin{equation*}
    i \frac{d \hat{a}_j^\dagger}{dz}= C (\hat{a}^{\dagger}_{j+1} + \hat{a}^{\dagger}_{j-1}) + \kappa \hat{a}^{\dagger}_{N+1}, ~~~(j=2,3,...,N-1) 
\end{equation*}
and
\begin{equation}
    i \frac{d \hat{a}_N^\dagger}{dz}= C (\hat{a}^{\dagger}_1 + \hat{a}^{\dagger}_{N-1}) + \kappa \hat{a}^{\dagger}_{N+1}.
\end{equation}
The solution, $\hat{a}_{N+1}^{\dagger}(z)$, can be written as 

\begin{equation*}
\hat{a}_{N+1}^\dagger(z)= e^{(-iCz)} \bigg\{ \bigg[ cos \big(\sqrt{C^2 + N\kappa^2}z \big) 
\end{equation*}
\begin{equation*}
+ \dfrac{iC}{\sqrt{C^2 + N \kappa^2}} sin \big(\sqrt{C^2 + N \kappa^2} z \big) \bigg]\hat{a}_{N+1}^{\dagger} (0)
\end{equation*}
\begin{equation}\label{eq12}
  - \dfrac{i \kappa}{\sqrt{C^2 + N \kappa^2}} sin \big(\sqrt{C^2 + N \kappa^2} z \big) \sum_{j=1}^{N} \hat{a}_j^{\dagger}(0) \bigg\}
\end{equation}

with $N \kappa^2 = C^2$, $\hat{a}_{N+1}^{\dagger}(z)$ is written as 
\begin{equation*}
\hat{a}_{N+1}^\dagger(z)= e^{(-iCz)} \bigg\{ \bigg[ cos \big(\sqrt{2}Cz \big)  + \dfrac{i}{\sqrt{2}} sin \big(\sqrt{2}C z \big) \bigg] \hat{a}_{N+1}^{\dagger} (0)    
\end{equation*}
\begin{equation}
    - \dfrac{i}{\sqrt{2N}} sin \big(\sqrt{2}C z \big) \sum_{j=1}^{N} \hat{a}_j^{\dagger}(0) \bigg\}.
\end{equation}

When $Cz = \dfrac{n\pi}{2\sqrt{2}}$, with $n$ being odd integer, the cosine term becomes zero. For $n=1$, the input state $\hat{a}^{\dagger}_{N+1} \vert 000....0 \rangle_{1...N+1}$ evolves to 
\begin{equation*}
 \vert W'\rangle_{N+1} = e^{i\pi/2 \sqrt{2}} \bigg[ \dfrac{i}{\sqrt{2N}} \bigg( \vert 10....0\rangle_{1...N}\vert 0 \rangle_{N+1} + ....
\end{equation*}
\begin{equation}
+ \vert 00....1\rangle_{1...N}\vert 0 \rangle_{N+1} \bigg)- \dfrac{i}{\sqrt{2}} \vert 00....0 \rangle_{1...N} \vert 1\rangle_{N+1} \bigg]     
\end{equation}

In the 2D ring structure, the radius $(r)$ of the ring and the nearest neighbor distance $(a)$ between surrounding waveguides are related as $a=2r sin(\pi/N)$. Due to the condition, $N\kappa^2 = C^2$, the value of $r$ and hence the value of $a$ depend on the number of surrounding waveguides $(N)$. This dependence can be found, by taking $\kappa = k e^{-r/d_0}$ and $C=k e^{-a/d_0}$ ($k$ is characteristic coupling strength and $d_0$ describes the rate of exponential decay of coupling strength), as 
\begin{equation}
\dfrac{r}{d_0} = \dfrac{ log_e(\sqrt{N})}{1-2 sin(\pi/N)}~~~ with~~~ N > 6
\end{equation}

\begin{figure}[h]
\centering
\includegraphics[scale=0.3]{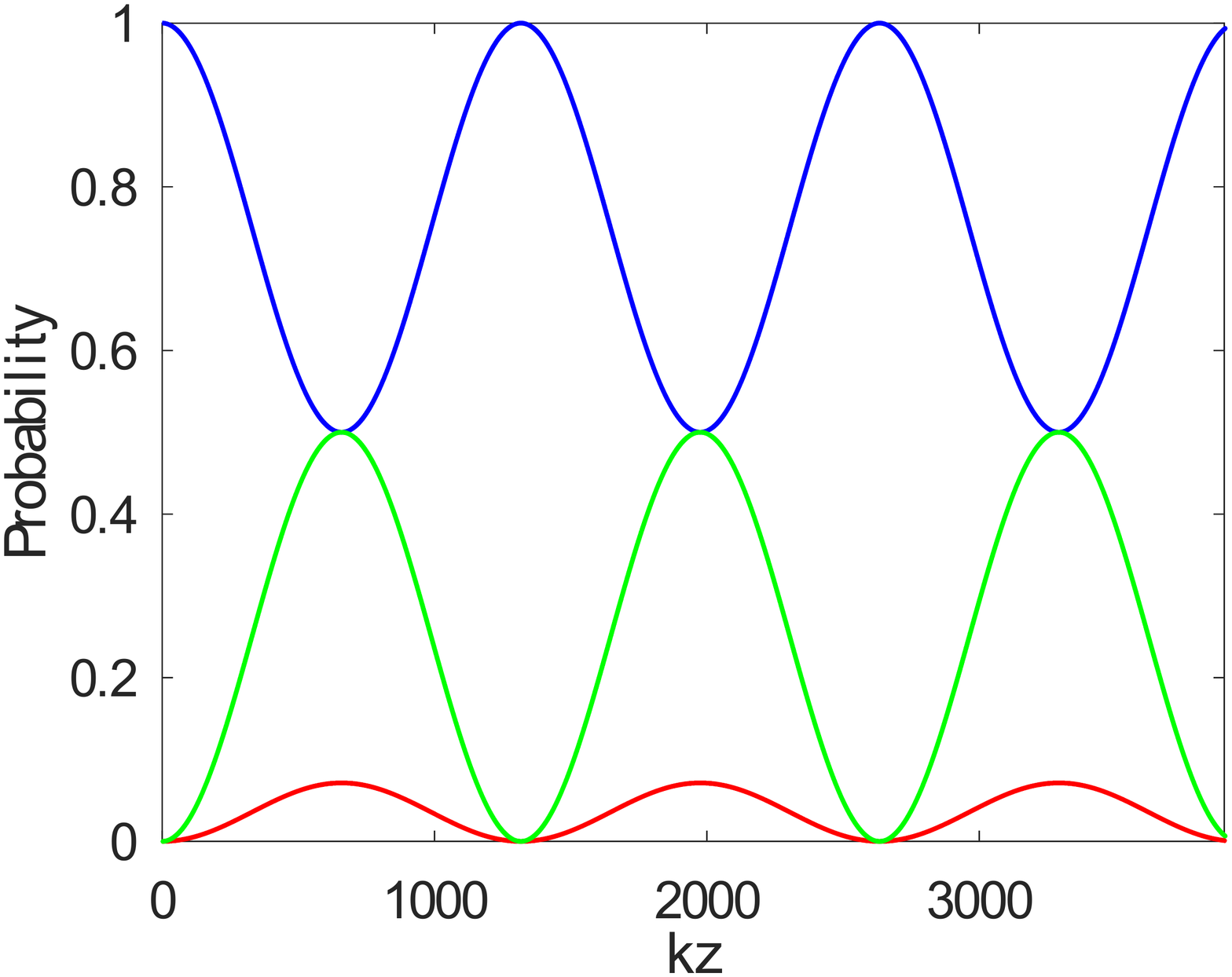}
\caption{Probabilities as function of $kz$ for 2D ring structure with 7 surrounding  waveguides. The blue (green) curve represents the probability of finding the photon at the central waveguide (any one of surrounding waveguides); red curve represents the probability of finding the photon in a specific surrounding waveguide which is same for all surrounding waveguides.}
\label{Fig.3}
\end{figure}

For example, when $N=7$, we have, $r \approx 7.35791 ~d_0$, $a \approx 6.38496 ~d_0$, $C \approx 1.6867 \times 10^{-3} ~k$ and $\kappa \approx 0.6375 \times 10^{-3} ~k$. The probabilities as function of $kz$ are shown in Figure \ref{Fig.3}. The blue and green curves represent the probabilities of finding the photon at the central waveguide and surrounding waveguides respectively. The red curve represents the probability of finding the photon at a specific surrounding waveguide. The values of $kz$ where the blue and green curves intersect, the 8-mode single photon perfect W-state can be generated by introducing appropriate phase shifts at individual waveguides. 

It can be verified that the ratio $a/d_0$ decreases as $N$ is increased. Hence, in this scheme, $N$ cannot have very large values. In addition, for $N > 12$, the second nearest neighbor distance between surrounding waveguides becomes smaller than $r$ which can be verified from the geometry of 2D ring structure (Figure \ref{Fig.2}) and hence the higher order coupling between surrounding waveguides are not ignorable. It can be noted that with the coupling scheme, $\kappa = C$, to generate single photon symmetric W-state~\cite{perez2013generating}, one can have only 6 surrounding waveguides. 

Further, it can be noted that the interaction part of the Hamiltonian (Eq. \ref{eq10}) with $C=0$ can be used to describe the intercavity hopping~\cite{meher2017duality} of a single photon in a system consisting of $N+1$ cavities in which $N$ cavities are coupled to a cavity through optical fibers. The fiber modes can be eliminated by adiabatic elimination process~\cite{pellizzari1997quantum,van1999quantum,lu2019geometrical,kyoseva2012coherent,cho2008heralded}. This system can be used to generate $N+1$- cavity mode single photon perfect W-state.  

Earlier, cavity QED schemes to prepare single photon cavity mode symmetric W-state was proposed~\cite{bergou1999entangled,guo2002scheme,yang2004scheme}. These schemes involve interaction of an atom in the excited state with multiple cavities in the vacuum state. In these schemes, atom-cavity field interaction time has to be precisely controlled to generate the desired entangled states. The lifetime of created entangled states and hence the implementation of any application is limited by the damping time of cavities. Hence these schemes require high $Q$ cavities. As compared to these cavity QED schemes, generation of single photon path entangled states using waveguide systems is more feasible. Generation of 16-mode single photon symmetric W-state using waveguides has been reported~\cite{grafe2014chip}. Hence the schemes that we have proposed using weakly coupled waveguide structures are more promosing to generate single photon path entangled perfect W-states.   

\section{Generalized entanglement detection conditions}
Generalized $N$-mode single photon perfect W-states belonging to W-class are not biseparable and retain pairwise entanglement. Hence the entanglement detection condition proposed in Ref.~\cite{selvan2019entanglement} can be used to detect the entanglement of generalized $N$-mode single photon perfect W-states. However, there are certain generalized $N$-mode single photon perfect W-states which will not satisfy that condition. For example, the generalized 5-mode single photon perfect W-state, $\vert \Tilde{W} \rangle_5 = \dfrac{1}{2\sqrt{2}} \big( \vert 10000 \rangle + \vert 01000 \rangle - \vert 00100 \rangle - \vert 00010 \rangle + 2 \vert 00001 \rangle \big)$, will not satisy the following 5-mode entanglement condition~\cite{selvan2019entanglement}.
\begin{equation}
    \vert \langle (\hat{a}_1+\hat{a}_2+\hat{a}_3+\hat{a}_4)\hat{a}_5^{\dagger} \rangle \vert^2 > 4 \langle \hat{N}_{\hat{a}_1+\hat{a}_2+\hat{a}_3+\hat{a}_4} \hat{N}_{\hat{a}_5} \rangle
\end{equation}

where $$\hat{N}_{\hat{a}_1 + \hat{a}_2 + \hat{a}_3 + \hat{a}_4} = \dfrac{(\hat{a}_1 + \hat{a}_2 + \hat{a}_3 + \hat{a}_4)^{\dagger} (\hat{a}_1 + \hat{a}_2 + \hat{a}_3 + \hat{a}_4)}{4}.$$

Hence, we derive a new set of entanglement conditions
suitable to detect the entanglement of generalized $N$-mode single photon perfect W-states. From the form of generalized $N$-mode perfect W-states [Eq.~\ref{eq1}], we consider the following operators.
 
\begin{equation*}
    \hat{L}_1 = \sum_{j=1}^{N-1} \big[ \alpha_j \hat{a}_j \hat{a}^{\dagger}_N + \alpha_j^* \hat{a}_j^{\dagger} \hat{a}_N \big] 
\end{equation*}

\begin{equation*}
   \hat{L}_2 = \sum_{j=1}^{N-1} \big[i \alpha_j \hat{a}_j \hat{a}^{\dagger}_N - i \alpha_j^* \hat{a}_j^{\dagger} \hat{a}_N \big] 
\end{equation*}
and
\begin{equation}\label{su}
    \hat{L}_3= \sum_{j,k=1}^{N-1} \big[ \alpha_j^* \alpha_k \hat{a}_j^{\dagger} \hat{a}_k \big] - \hat{N}_{\hat{a}_N}
\end{equation}

 where $\alpha_j$'s are non-zero complex coefficients as described in the Introduction. For the given set of $\alpha_j$'s, these three operators satisfy $su(2)$ algebra, $\big[ \hat{L}_x, \hat{L}_y \big] = 2i \epsilon_{xyz} \hat{L}_z$ with $x,y,z=1,2,3$. The sum of variances of $\hat{L}_1$ and $\hat{L}_2$ can be written as 
 
 \begin{equation*}
   \big(\Delta \hat{L}_1 \big)^2 + \big( \Delta \hat{L}_2 \big)^2 =  2 \bigg [\sum_{j,k=1}^{N-1} \langle \alpha_j^* \alpha_k \hat{a}_j^{\dagger} \hat{a}_k \rangle + \langle \hat{N}_{\hat{a}_N} \rangle \bigg]  
 \end{equation*}
 \begin{equation}
      + 4 \bigg[ \sum_{j,k=1}^{N-1} \langle \alpha_j^* \alpha_k \hat{a}_j^{\dagger} \hat{a}_k \hat{N}_{\hat{a}_N} \rangle - \bigg| \sum_{j=1}^{N-1} \langle \alpha_j \hat{a}_j \hat{a}_N^{\dagger} \rangle \bigg|^2 \bigg]
 \end{equation}
 
 Following the arguments given in Ref.~\cite{selvan2019entanglement}, it can be shown that the following inequality relation is satisfied by fully separable states. 
 
 \begin{equation}\label{ed3}
    (\Delta \hat{L}_1)^2 + (\Delta \hat{L}_2)^2 \geq 2 \bigg[ \sum_{j,k=1}^{N-1}  \langle  \alpha_j^* \alpha_k \hat{a}_j^{\dagger} \hat{a}_{k}\rangle  + \langle \hat{N}_{\hat{a}_N} \rangle \bigg]
\end{equation}

Hence, violation of this inequality relation implies that the state is entangled and the entanglement condition can be written as 

\begin{equation}\label{ed4}
   \bigg \vert \sum_{j=1}^{N-1} \langle \alpha_j \hat{a}_j \hat{a}_N^{\dagger} \rangle \bigg \vert^2 > \sum_{j,k=1}^{N-1}  \langle \alpha_j^* \alpha_k\hat{a}_j^{\dagger} \hat{a}_{k} \hat{N}_{\hat{a}_N} \rangle
\end{equation}

The entanglement condition proposed in Ref.~\cite{selvan2019entanglement} is a special case $(\alpha_j = {1}/{\sqrt{N-1}},~~\forall j)$ of this condition. Genuine entanglement of multimode states which retain pairwise entanglement between all possible pairs of modes, referred as multimode W-type entangled states in Ref.~\cite{selvan2019entanglement}, can be detected using this set of conditions. These states include both single photon and multiphoton states. Furthermore, it is possible to obtain higher order entanglement conditions, as discussed in Refs.~\cite{hillery2006entanglement,selvan2019entanglement}, to detect the genuine entanglement of more W-type multiphoton entangled states.

Since the $su(2)$ algebra operators~(Eq.\ref{su}) conserve the total number of photons, we now consider the action of these operators only on the single photon subspace. Any genuinely entangled $N$-mode single photon state of the form 
\begin{equation}\label{eq26}
    \vert \Psi \rangle = \sum_{j=1}^N C_j \vert 1 \rangle_j \vert 0...0 \rangle_{1..j-1,j+1..N},
\end{equation}

with $C_j = \vert C_j \vert e^{i\varphi_j}$ and $\sum_{j=1}^{N}\vert C_j \vert^2 = 1$, can be rewritten in two different ways as (global phase term $e^{i\varphi_N}$ is excluded)

\begin{equation*}
\vert \Psi \rangle = \pm \sqrt{\dfrac{1+\lambda}{2}} \bigg[\sum_{j=1}^{N-1} \alpha_j^*  \vert 1 \rangle_{j} \vert 0...0\rangle_{1..j-1,j+1..N} \bigg]   
\end{equation*}
\begin{equation}
+ \sqrt{\dfrac{1-\lambda}{2}} \vert 00....01 \rangle_{1...N}
\end{equation}

where

\begin{equation}\label{eq26a}
    \lambda = 1 - 2 \vert C_N \vert^2~~\text{and}~~\alpha_j^*= \dfrac{ \pm C_j e^{-i\varphi_N}}{\sqrt{\sum_{j=1}^{N-1}\vert C_j \vert^2}}.
\end{equation}

Since $\vert C_N \vert$ is neither 1 nor 0, the value of $\vert \lambda \vert$ is less than one. 

The state, $\vert \Psi \rangle$ can be shown to be $\hat{L}_1$-squeezed state. That is, it can be shown to satisfy the following conditions~\cite{nha2006entanglement,hillery1993interferometers}. 

\begin{equation*}
   \big( \Delta \hat{L}_1 \big) \big( \Delta \hat{L}_2 \big) = \vert \langle \hat{L}_3 \rangle \vert
\end{equation*}

\begin{equation*}
    \big(\Delta \hat{L}_1 \big)^2 = \vert \lambda \langle \hat{L}_3 \rangle \vert
\end{equation*}
and
\begin{equation}
    \big(\Delta \hat{L}_2 \big)^2 = \bigg| \dfrac{\langle \hat{L}_3 \rangle}{\lambda} \bigg|
\end{equation}

Further, the state $\vert \Psi \rangle$ can be written as superposition of two orthonormal generalized $N$-mode single photon perfect W-states. That is, 

\begin{equation*}
    \vert \Psi \rangle = D_1 \vert \Tilde{W} \rangle_{N+} + D_2 \vert \Tilde{W} \rangle_{N-}
\end{equation*}
with
\begin{equation*}
    D_1 = \dfrac{\sqrt{1-\lambda} \pm \sqrt{1+\lambda}}{2}, 
\end{equation*}
\begin{equation*}
    D_2 = \dfrac{\sqrt{1-\lambda} \mp \sqrt{1+\lambda}}{2}, 
\end{equation*}
and 
\begin{equation*}
  \vert \Tilde{W} \rangle_{N\pm} = \pm \dfrac{1}{\sqrt{2}} \bigg[\sum_{j=1}^{N-1} \alpha_j^*  \vert 1 \rangle_{j} \vert 0...0\rangle_{1..j-1,j+1..N} \bigg]  
\end{equation*}
\begin{equation}{\label{eig}}
    + \dfrac{1}{\sqrt{2}} \vert 00....01 \rangle_{1...N}
\end{equation}

It can be verified that the states $\vert \Tilde{W} \rangle_{N\pm}$ are eigenstates of $\hat{L}_1$ corresponding to the eigenvalues $\pm 1$. 

The $\hat{L}_2$-squeezed states corresponding to the $\hat{L}_1$-squeezed states given in Eq.~\ref{eq26} can be written as 

\begin{equation}
    \vert \Psi'\rangle = -i\sum_{j=1}^{N-1} C_j \vert 1 \rangle_j \vert 0...0 \rangle_{1..j-1,j+1..N} + C_N \vert 00...01 \rangle_{1...N}
\end{equation}

The relation between $\alpha_j$'s and $C_j$'s, and the squeezing parameter $\lambda$ and $C_N$ are the same as given in Eq.~\ref{eq26a}. Similar to $\hat{L}_1$-squeezed states, the $\hat{L}_2$-squeezed states can be expressed as linear combination of two orthonormal generalized $N$-mode single photon perfect W-states as follows. 

\begin{equation}
    \vert \Psi' \rangle = D_1 \vert W'\rangle_{+} + D_2 \vert W'\rangle_{-}
\end{equation}

where, 

\begin{equation}
  \vert {W'} \rangle_{N\pm} = \mp i \dfrac{1}{\sqrt{2}} \bigg[ \sum_{j=1}^{N-1} \alpha_j^*  \vert 1 \rangle_{j} \vert 0...0\rangle_{1..j-1,j+1..N} \bigg]  
\end{equation}
\begin{equation}
      + \dfrac{1}{\sqrt{2}} \vert 00....01 \rangle_{1...N}
\end{equation}

are eigenstates of $\hat{L}_2$ corrseponding to the eigenvalues $\pm1$. 

Further the action of operators involved in the separability condition [Eq.~\ref{ed3}], within the single photon subspace, can be represented by the action of the projection operators as shown below.

\begin{equation}
    \hat{L}_1 \vert \Phi \rangle = \bigg[\vert \tilde{W} \rangle_{{N+}} {_{+N}} \langle \tilde{W} \vert - \vert \tilde{W} \rangle_{N-} {_{-N}} \langle \tilde{W} \vert \bigg] \vert \Phi \rangle
\end{equation}

\begin{equation}
    \hat{L}_2 \vert \Phi \rangle = \bigg[ \vert {W'} \rangle_{{N+}} {_{+N}} \langle {'W} \vert - \vert {W'} \rangle_{N-} {_{-N}} \langle {'W} \vert \bigg] \vert \Phi \rangle 
\end{equation}

\begin{equation*}
  \bigg[ \sum_{j,k=1}^{N-1}   \alpha_j^* \alpha_k \hat{a}_j^{\dagger} \hat{a}_{k}  + \hat{N}_{\hat{a}_N} \bigg] \vert \Phi \rangle ~~~~~~~~~~~~~~~~~~~~~~~~~~~~~~~~~~~~~~~~~~
\end{equation*}
\begin{equation*}
  =\bigg[ \vert \tilde{W} \rangle_{{N+}} {_{+N}} \langle \tilde{W} \vert + \vert \tilde{W} \rangle_{N-} {_{-N}} \langle \tilde{W} \vert \bigg] \vert \Phi \rangle   
\end{equation*}
\begin{equation}
    = \bigg[ \vert {W'} \rangle_{{N+}} {_{+N}} \langle {'W} \vert + \vert {W'} \rangle_{N-} {_{-N}} \langle {'W} \vert \bigg] \vert \Phi \rangle
\end{equation}

where $\vert \Phi \rangle$ is generated $N$-mode single photon state. 

In terms of these projection operators, the separability condition given in Eq.~\ref{ed3} can be rewritten as 

\begin{equation*}
\bigg[ \big| \langle \Phi \vert \tilde{W} \rangle_{N+} \big|^2 - \big| \langle \Phi \vert \tilde{W} \rangle_{N-} \big|^2 \bigg]^2 ~~~~~~~~~~~~~~~~~~~~~~~~~~~~~~~~~~~~~~~~~~~
\end{equation*}
\begin{equation}\label{fied}
 + \bigg[ \big| \langle \Phi \vert W' \rangle_{N+} \big|^2 - \big| \langle \Phi \vert W' \rangle_{N-} \big|^2 \bigg]^2 \leq 0
\end{equation}

We refer this condition as single photon separability condition. The two terms on the left hand side (LHS) of single photon separability condition are always positive and hence the terms on the LHS being not equal to zero indicate that the single photon state, $\vert \Phi \rangle$, is entangled. It can be verifed that for $\hat{L}_1 (\hat{L}_2)$-squeezed state, $\vert \Psi \rangle$ ($\vert \Psi' \rangle$), the second (first) term on the LHS of single photon separability condition becomes zero and that condition becomes

\begin{equation}
    1- \lambda^2 \leq 0. 
\end{equation}

With $\lambda \in (-1,1)$, the single photon separability condition is violated by $\hat{L}_1$ and $\hat{L}_2$-squeezed states. Hence, in order to verify the entanglement of any given genuinely entangled $N$-mode single photon state, one has to find the difference between the fidelities of the given $N$-mode single photon state with properly chosen set of two orthonormal generalized $N$-mode single photon perfect W-states. 

\section{Experimental implementation of generalized entanglement detection conditions}

\begin{figure*}[htp]
\centering
\includegraphics[scale=1.5]{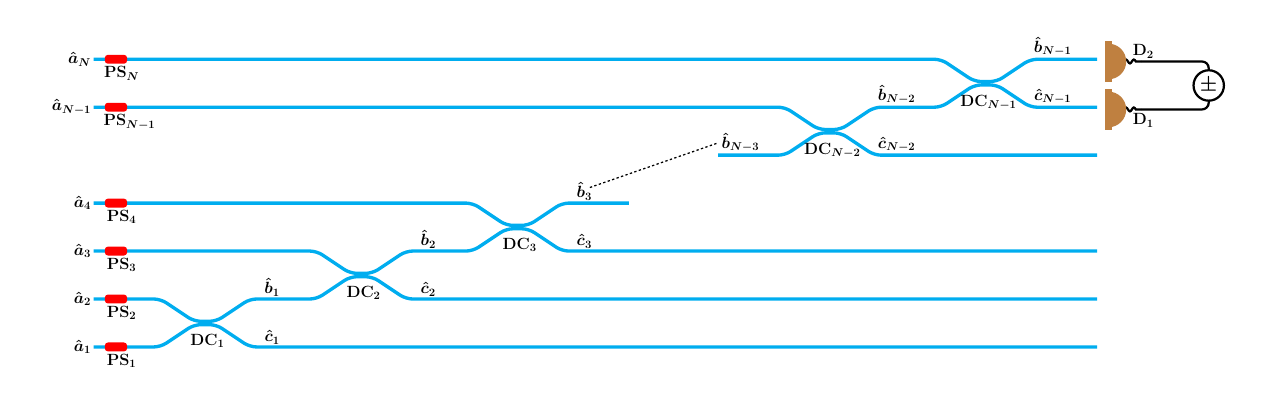}
\caption{Integrated photonic circuit to verify the violation of separability conditions (Eq.~\ref{ed3} and~\ref{fied}). It has $(N - 1)$ directional couplers (DC$_j$'s) and $N$ phase shifters (PS$_j$'s) as described in the main text. The dashed line represents the repetition of the pattern of circuit from DC$_3$ to DC$_{N-2}$. D$_1$ and D$_2$ are photodetectors.}
\label{Fig.4}
\end{figure*}

We now consider the experimental verification of violation of separability condition proposed in the last section. In order to do that the operators $\hat{L}_1$, $\hat{L}_2$ and the one on the right hand side of Eq.~\ref{ed3} have to be measured. An integrated photonic circuit to measure these operators is shown in Figure \ref{Fig.4}. The circuit consists of $N-1$ directional couplers and $N$ phase shifters. This circuit can be fabricated on a chip. 

The action of directional coupler DC$_j$ on the input modes $\hat{b}_{j-1}$ and $\hat{a}_{j+1}$ is shown below.
\begin{equation}\label{exp1}
    \begin{pmatrix}
     \hat{b}_j \\ \hat{c}_j
    \end{pmatrix} 
    = 
    \begin{pmatrix}
     T_j & R_j \\ R_j & T_j 
    \end{pmatrix} 
    \begin{pmatrix}
     \hat{b}_{j-1} \\ \hat{a}_{j+1}
    \end{pmatrix}
\end{equation}

where, $T_j$ and $R_j$ satisfy the conditions, $\vert T_j \vert^2 + \vert R_j \vert^2 = 1$ and $R_j^*T_j + T_j^*R_j = 0$. 

The input modes of directional coupler DC$_1$ are $\hat{a}_1 =(\hat{b}_0)$ and $\hat{a}_2$. The action of phase shifter PS$_j$ on the mode $\hat{a}_j$ is given by

\begin{equation}\label{exp2}
    \hat{a}_j \rightarrow \hat{a}_j e^{-i \phi_j}
\end{equation}

From Eqs.~\ref{exp1} and \ref{exp2}, the mode $\hat{b}_{N-2}$ can be written as follows.

\begin{equation}\label{exp3}
\hat{b}_{N-2}= \sum_{j=1}^{N-1} \alpha_j \hat{a}_j
\end{equation}

where

\begin{equation*}
    \alpha_1 = \bigg[ \prod_{k=1}^{N-2} T_{N-1-k} \bigg] e^{-i \phi_1},
\end{equation*}

\begin{equation*}
    \alpha_j = \bigg[ \prod_{k=1}^{N-1-j}T_{N-1-k} \bigg] R_{j-1}e^{-i \phi_j},~~~~~~~~j=2,...,N-2
\end{equation*}
and 
\begin{equation}\label{exp4}
\alpha_{N-1} = R_{N-2} e^{-i \phi_{N-1}}.
\end{equation}

It can be verified that $\sum_{j=1}^{N-1} \vert \alpha_j \vert^2 = 1$. 

For directional coupler, DC$_{N-1}$, the values of $R_{N-1}$ and $T_{N-1}$ are chosen as $i/\sqrt{2}$ and $1/\sqrt{2}$ respectively. The phase shift introduced by the phase shifter PS$_N$ is denoted as $\phi_N$. The output modes $\hat{b}_{N-1}$ and $\hat{c}_{N-1}$ of DC$_{N-1}$ can be written as 
\begin{equation}\label{eq31}
\hat{b}_{N-1} = \dfrac{\hat{b}_{N-2} + i\hat{a}_N e^{-i\phi_N}}{\sqrt{2}}
\end{equation}

\begin{equation}\label{eq32}
\hat{c}_{N-1} = \dfrac{i\hat{b}_{N-2} + \hat{a}_N e^{-i\phi_N}}{\sqrt{2}}
\end{equation}

For $\phi_N=\pi/2$ and $\pi$, the photon number difference $\hat{b}_{N-1}^{\dagger}\hat{b}_{N-1} - \hat{c}_{N-1}^{\dagger}\hat{c}_{N-1}$ gives the operators $\hat{L}_1$ and $\hat{L}_2$ respectively. Similarly, $\hat{b}_{N-1}^{\dagger}\hat{b}_{N-1} + \hat{c}_{N-1}^{\dagger}\hat{c}_{N-1}$ yields the operator on the right hand side of Eq.~\ref{ed3} (excluding the factor 2) for any value of $\phi_N$. Thus by measuring the sum and difference of photon numbers at photodetectors D$_1$ and D$_2$, the violation of separability condition can be verified.

Measurement of the two terms involved in the single photon separability condition [Eq.~\ref{fied}] can be explained as follows. When the input state is $\vert \Tilde{W} \rangle_{N+}$ ($\vert \Tilde{W} \rangle_{N-}$), the photonic circuit with $\phi_N = \pi/2$ does a unitary transformation such that the photon will be found in the mode $\hat{b}_{N-1}$ ($\hat{c}_{N-1}$) at the output. This can be verified from Eq.~\ref{eq31} (Eq.~\ref{eq32}). When the input state is one of the remaining $N-2$ degenerate eigenstates of $\hat{L}_1$ corresponding to the eigenvalue $0$, the photon will be found in one of the remaining $N-2$ output modes which can be verified from the input-output relations of this circuit. The transformation of generated $N$-mode single photon state, $\vert \Phi \rangle$, can be found by expressing the state in the eigenbasis of $\hat{L}_1$. Thus measuring the photon number difference at the photodetectors D$_1$ and D$_2$ gives the square root of the first term of Eq.~\ref{fied}. Similarly, the measurement of second term can be explained by choosing $\phi_N = \pi$ and expressing the state, $\vert \Phi \rangle$, in the eigenbasis of $\hat{L}_2$. 

The values of $R_j$'s, $T_j$'s ($j=1,2,...,N-2$) and $\phi_k$'s ($k=1,2,...,N-1$) can be chosen suitably to detect the entanglement of a given generalized $N$-mode single photon perfect W-state. For example, in the case of $N$-mode single photon perfect W-state [Eq.~\ref{eq2}], the values of $R_j$ and $T_j$ ($j=1,2,3,...,N-2$) have to be chosen as $i/\sqrt{j+1}$ and $\sqrt{j/(j+1)}$ respectively. The phase shift $(\phi_j)$ introduced by the phase shifter PS$_j$ for $j=2,3,...N-1$ have to be $\pi/2$ and $0$ for $j=1$.

As compared to the single photon entanglement detection scheme presented in Ref.~\cite{lougovski2009verifying}, our scheme involves very much reduced measurement setup. In our scheme, in order to detect the entanglement of genuinely entangled single photon states, the photon has to be detected only at two output modes irrespective of the number of input modes. It is also possible to experimentally characterize the unitary transformation performed by the photonic circuit (Figure~\ref{Fig.4}) as described in Ref.~\cite{heilmann2015novel}. In the earlier experimental work~\cite{grafe2014chip}, the events due to imperfections, discussed in Ref.~\cite{lougovski2009verifying}, were shown to have negligible influence on the generation and verification of single photon entangled states using waveguide structures, single photon source based on spontaneous parametric down-conversion and avalanche photodiodes. Hence, with reduced measurement setup, our entanglement detection scheme can easily be implemented and is more suitable for detecting the entanglement of genuinely entangled single photon states with large number of modes. It can be noted that the application of entanglement condition given in Eq.~\ref{ed4}, goes beyond single photon entangled states as mentioned in the last section. 

Finally, it can be verified from Eqs.~\ref{eq31} and \ref{eq32} that the generalized $N$-mode single photon perfect W-states can be generated, using the photonic circuit given in Figure~\ref{Fig.4}, by injecting a photon in either the mode $\hat{b}_{N-1}$ or $\hat{c}_{N-1}$.

\section{Conclusion}
We have proposed generation and entanglement verification schemes for generalized $N$-mode single photon perfect W-states. The single photon realization of generalized $N$-qubit perfect W-states are more advantageous than the other realizations~\cite{rao2008generation,li2016generating,zhao2008scheme,swain2020generation} as the single photon states can be generated easily using waveguide structures. These structures are stable, scalable, and offer low photon propagation losses~\cite{chen2017exact,chen2018entanglement}. The state generation schemes involve injecting a photon into one of $N$ weakly coupled waveguides arranged in two different geometrical structures: 1D planar and 2D ring structures. 1D planar structure is scalable to generate large $N$-mode single photon perfect W-states whereas the 2D ring structure has restrictions on the number of surrounding waveguides. 

The generalized entanglement detection conditions are obtained from sum uncertainty relations of generalized set of $su(2)$ algebra operators. They are suitable to detect the entanglement of multimode W-type entangled states~\cite{selvan2019entanglement}. The entanglement verification of generalized $N$-mode single photon perfect W-states can be performed using the single photon separability conditions which are obtained by diagonalizing the operators, involved in the generalized entanglement conditions, within the single photon subspace. As the generalized $N$-mode single photon perfect W-states are single photon eigenstates of $su(2)$ algebra operators corresponding to non-zero eigenvalues, verification of entanglement of single photon states involves finding fidelity with generalized $N$-mode perfect W-states. The genuinely entangled single photon states are squeezed states of suitably chosen set of $su(2)$ algebra operators and they violate the corresponding single photon separability condition. Experimental implementation of entanglement conditions require doing measurement at two specific outputs of a photonic circuit consisting of directional couplers and phase shifters, irrespective of number of inputs. Thus the proposed entanglement conditions are more suitable not only for entanglement verification of generalized $N$-mode single photon perfect W-states but also to detect the entanglement of any genuinely entangled $N$-mode single photon states with large $N$. 

The integrated photonic circuit used to implement the entanglement detection conditions can also be used to generate generalized $N$-mode single photon perfect W-states.  To the best of our knowledge, the generalized $N$-qubit perfect W-states with $N > 3$ are not generated experimentally. Hence, the waveguide structures are more promising candidates to generate generalized $N$-mode single photon perfect states with large $N$ than bulk optical elements~\cite{li2016generating} and superconducting qubits~\cite{swain2020generation} as the waveguide structures are already proved to be suitable to generate maximally entangled single photon states~\cite{grafe2014chip}.

\section*{Acknowledgments}
A.R gratefully acknowledges a research grant from Science
and Engineering Research Board (SERB), Department of Science and Technology (DST), Government of India (Grant No.:CRG/2019/005749) during this work. P. K. P acknowledges a research grant from SERB, DST project  (Grant No.:DST/ICPS/QuST/Theme-1/2019/2020-21/01).

\bibliography{main}

\begin{thebibliography}{77}%
\makeatletter
\providecommand \@ifxundefined [1]{%
 \@ifx{#1\undefined}
}%
\providecommand \@ifnum [1]{%
 \ifnum #1\expandafter \@firstoftwo
 \else \expandafter \@secondoftwo
 \fi
}%
\providecommand \@ifx [1]{%
 \ifx #1\expandafter \@firstoftwo
 \else \expandafter \@secondoftwo
 \fi
}%
\providecommand \natexlab [1]{#1}%
\providecommand \enquote  [1]{``#1''}%
\providecommand \bibnamefont  [1]{#1}%
\providecommand \bibfnamefont [1]{#1}%
\providecommand \citenamefont [1]{#1}%
\providecommand \href@noop [0]{\@secondoftwo}%
\providecommand \href [0]{\begingroup \@sanitize@url \@href}%
\providecommand \@href[1]{\@@startlink{#1}\@@href}%
\providecommand \@@href[1]{\endgroup#1\@@endlink}%
\providecommand \@sanitize@url [0]{\catcode `\\12\catcode `\$12\catcode
  `\&12\catcode `\#12\catcode `\^12\catcode `\_12\catcode `\%12\relax}%
\providecommand \@@startlink[1]{}%
\providecommand \@@endlink[0]{}%
\providecommand \url  [0]{\begingroup\@sanitize@url \@url }%
\providecommand \@url [1]{\endgroup\@href {#1}{\urlprefix }}%
\providecommand \urlprefix  [0]{URL }%
\providecommand \Eprint [0]{\href }%
\providecommand \doibase [0]{https://doi.org/}%
\providecommand \selectlanguage [0]{\@gobble}%
\providecommand \bibinfo  [0]{\@secondoftwo}%
\providecommand \bibfield  [0]{\@secondoftwo}%
\providecommand \translation [1]{[#1]}%
\providecommand \BibitemOpen [0]{}%
\providecommand \bibitemStop [0]{}%
\providecommand \bibitemNoStop [0]{.\EOS\space}%
\providecommand \EOS [0]{\spacefactor3000\relax}%
\providecommand \BibitemShut  [1]{\csname bibitem#1\endcsname}%
\let\auto@bib@innerbib\@empty
\bibitem [{\citenamefont {Borras}\ \emph {et~al.}(2007)\citenamefont {Borras},
  \citenamefont {Plastino}, \citenamefont {Batle}, \citenamefont {Zander},
  \citenamefont {Casas},\ and\ \citenamefont
  {Plastino}}]{borras2007multiqubit}%
  \BibitemOpen
  \bibfield  {author} {\bibinfo {author} {\bibfnamefont {A.}~\bibnamefont
  {Borras}}, \bibinfo {author} {\bibfnamefont {A.}~\bibnamefont {Plastino}},
  \bibinfo {author} {\bibfnamefont {J.}~\bibnamefont {Batle}}, \bibinfo
  {author} {\bibfnamefont {C.}~\bibnamefont {Zander}}, \bibinfo {author}
  {\bibfnamefont {M.}~\bibnamefont {Casas}},\ and\ \bibinfo {author}
  {\bibfnamefont {A.}~\bibnamefont {Plastino}},\ }\bibfield  {title} {\bibinfo
  {title} {Multiqubit systems: highly entangled states and entanglement
  distribution},\ }\href@noop {} {\bibfield  {journal} {\bibinfo  {journal}
  {Journal of Physics A: Mathematical and Theoretical}\ }\textbf {\bibinfo
  {volume} {40}},\ \bibinfo {pages} {13407} (\bibinfo {year}
  {2007})}\BibitemShut {NoStop}%
\bibitem [{\citenamefont {Facchi}\ \emph {et~al.}(2008)\citenamefont {Facchi},
  \citenamefont {Florio}, \citenamefont {Parisi},\ and\ \citenamefont
  {Pascazio}}]{facchi2008maximally}%
  \BibitemOpen
  \bibfield  {author} {\bibinfo {author} {\bibfnamefont {P.}~\bibnamefont
  {Facchi}}, \bibinfo {author} {\bibfnamefont {G.}~\bibnamefont {Florio}},
  \bibinfo {author} {\bibfnamefont {G.}~\bibnamefont {Parisi}},\ and\ \bibinfo
  {author} {\bibfnamefont {S.}~\bibnamefont {Pascazio}},\ }\bibfield  {title}
  {\bibinfo {title} {Maximally multipartite entangled states},\ }\href@noop {}
  {\bibfield  {journal} {\bibinfo  {journal} {Physical Review A}\ }\textbf
  {\bibinfo {volume} {77}},\ \bibinfo {pages} {060304} (\bibinfo {year}
  {2008})}\BibitemShut {NoStop}%
\bibitem [{\citenamefont {Enr{\'\i}quez}\ \emph {et~al.}(2016)\citenamefont
  {Enr{\'\i}quez}, \citenamefont {Wintrowicz},\ and\ \citenamefont
  {{\.Z}yczkowski}}]{enriquez2016maximally}%
  \BibitemOpen
  \bibfield  {author} {\bibinfo {author} {\bibfnamefont {M.}~\bibnamefont
  {Enr{\'\i}quez}}, \bibinfo {author} {\bibfnamefont {I.}~\bibnamefont
  {Wintrowicz}},\ and\ \bibinfo {author} {\bibfnamefont {K.}~\bibnamefont
  {{\.Z}yczkowski}},\ }\bibfield  {title} {\bibinfo {title} {Maximally
  entangled multipartite states: a brief survey},\ }in\ \href@noop {} {\emph
  {\bibinfo {booktitle} {Journal of Physics: Conference Series}}},\ Vol.\
  \bibinfo {volume} {698}\ (\bibinfo {organization} {IOP Publishing},\ \bibinfo
  {year} {2016})\ p.\ \bibinfo {pages} {012003}\BibitemShut {NoStop}%
\bibitem [{\citenamefont {Miyake}(2003)}]{miyake2003classification}%
  \BibitemOpen
  \bibfield  {author} {\bibinfo {author} {\bibfnamefont {A.}~\bibnamefont
  {Miyake}},\ }\bibfield  {title} {\bibinfo {title} {Classification of
  multipartite entangled states by multidimensional determinants},\ }\href@noop
  {} {\bibfield  {journal} {\bibinfo  {journal} {Physical Review A}\ }\textbf
  {\bibinfo {volume} {67}},\ \bibinfo {pages} {012108} (\bibinfo {year}
  {2003})}\BibitemShut {NoStop}%
\bibitem [{\citenamefont {Koashi}\ \emph {et~al.}(2000)\citenamefont {Koashi},
  \citenamefont {Bu{\v{z}}ek},\ and\ \citenamefont
  {Imoto}}]{koashi2000entangled}%
  \BibitemOpen
  \bibfield  {author} {\bibinfo {author} {\bibfnamefont {M.}~\bibnamefont
  {Koashi}}, \bibinfo {author} {\bibfnamefont {V.}~\bibnamefont
  {Bu{\v{z}}ek}},\ and\ \bibinfo {author} {\bibfnamefont {N.}~\bibnamefont
  {Imoto}},\ }\bibfield  {title} {\bibinfo {title} {Entangled webs: Tight bound
  for symmetric sharing of entanglement},\ }\href@noop {} {\bibfield  {journal}
  {\bibinfo  {journal} {Physical Review A}\ }\textbf {\bibinfo {volume} {62}},\
  \bibinfo {pages} {050302} (\bibinfo {year} {2000})}\BibitemShut {NoStop}%
\bibitem [{\citenamefont {Kafatos}(2013)}]{kafatos2013bell}%
  \BibitemOpen
  \bibfield  {author} {\bibinfo {author} {\bibfnamefont {M.}~\bibnamefont
  {Kafatos}},\ }\href@noop {} {\emph {\bibinfo {title} {Bell's theorem, quantum
  theory and conceptions of the universe}}},\ Vol.~\bibinfo {volume} {37}\
  (\bibinfo  {publisher} {Springer Science \& Business Media},\ \bibinfo {year}
  {2013})\BibitemShut {NoStop}%
\bibitem [{\citenamefont {D{\"u}r}\ \emph {et~al.}(2000)\citenamefont
  {D{\"u}r}, \citenamefont {Vidal},\ and\ \citenamefont
  {Cirac}}]{dur2000three}%
  \BibitemOpen
  \bibfield  {author} {\bibinfo {author} {\bibfnamefont {W.}~\bibnamefont
  {D{\"u}r}}, \bibinfo {author} {\bibfnamefont {G.}~\bibnamefont {Vidal}},\
  and\ \bibinfo {author} {\bibfnamefont {J.~I.}\ \bibnamefont {Cirac}},\
  }\bibfield  {title} {\bibinfo {title} {Three qubits can be entangled in two
  inequivalent ways},\ }\href@noop {} {\bibfield  {journal} {\bibinfo
  {journal} {Physical Review A}\ }\textbf {\bibinfo {volume} {62}},\ \bibinfo
  {pages} {062314} (\bibinfo {year} {2000})}\BibitemShut {NoStop}%
\bibitem [{\citenamefont {D{\"u}r}\ \emph {et~al.}(2003)\citenamefont
  {D{\"u}r}, \citenamefont {Aschauer},\ and\ \citenamefont
  {Briegel}}]{dur2003multiparticle}%
  \BibitemOpen
  \bibfield  {author} {\bibinfo {author} {\bibfnamefont {W.}~\bibnamefont
  {D{\"u}r}}, \bibinfo {author} {\bibfnamefont {H.}~\bibnamefont {Aschauer}},\
  and\ \bibinfo {author} {\bibfnamefont {H.-J.}\ \bibnamefont {Briegel}},\
  }\bibfield  {title} {\bibinfo {title} {Multiparticle entanglement
  purification for graph states},\ }\href@noop {} {\bibfield  {journal}
  {\bibinfo  {journal} {Physical review letters}\ }\textbf {\bibinfo {volume}
  {91}},\ \bibinfo {pages} {107903} (\bibinfo {year} {2003})}\BibitemShut
  {NoStop}%
\bibitem [{\citenamefont {Hein}\ \emph {et~al.}(2004)\citenamefont {Hein},
  \citenamefont {Eisert},\ and\ \citenamefont {Briegel}}]{hein2004multiparty}%
  \BibitemOpen
  \bibfield  {author} {\bibinfo {author} {\bibfnamefont {M.}~\bibnamefont
  {Hein}}, \bibinfo {author} {\bibfnamefont {J.}~\bibnamefont {Eisert}},\ and\
  \bibinfo {author} {\bibfnamefont {H.~J.}\ \bibnamefont {Briegel}},\
  }\bibfield  {title} {\bibinfo {title} {Multiparty entanglement in graph
  states},\ }\href@noop {} {\bibfield  {journal} {\bibinfo  {journal} {Physical
  Review A}\ }\textbf {\bibinfo {volume} {69}},\ \bibinfo {pages} {062311}
  (\bibinfo {year} {2004})}\BibitemShut {NoStop}%
\bibitem [{\citenamefont {Nielsen}(2006)}]{nielsen2006cluster}%
  \BibitemOpen
  \bibfield  {author} {\bibinfo {author} {\bibfnamefont {M.~A.}\ \bibnamefont
  {Nielsen}},\ }\bibfield  {title} {\bibinfo {title} {Cluster-state quantum
  computation},\ }\href@noop {} {\bibfield  {journal} {\bibinfo  {journal}
  {Reports on Mathematical Physics}\ }\textbf {\bibinfo {volume} {57}},\
  \bibinfo {pages} {147} (\bibinfo {year} {2006})}\BibitemShut {NoStop}%
\bibitem [{\citenamefont {Kempe}(1999)}]{kempe1999multiparticle}%
  \BibitemOpen
  \bibfield  {author} {\bibinfo {author} {\bibfnamefont {J.}~\bibnamefont
  {Kempe}},\ }\bibfield  {title} {\bibinfo {title} {Multiparticle entanglement
  and its applications to cryptography},\ }\href@noop {} {\bibfield  {journal}
  {\bibinfo  {journal} {Physical Review A}\ }\textbf {\bibinfo {volume} {60}},\
  \bibinfo {pages} {910} (\bibinfo {year} {1999})}\BibitemShut {NoStop}%
\bibitem [{\citenamefont {Yeo}\ and\ \citenamefont
  {Chua}(2006)}]{yeo2006teleportation}%
  \BibitemOpen
  \bibfield  {author} {\bibinfo {author} {\bibfnamefont {Y.}~\bibnamefont
  {Yeo}}\ and\ \bibinfo {author} {\bibfnamefont {W.~K.}\ \bibnamefont {Chua}},\
  }\bibfield  {title} {\bibinfo {title} {Teleportation and dense coding with
  genuine multipartite entanglement},\ }\href@noop {} {\bibfield  {journal}
  {\bibinfo  {journal} {Physical review letters}\ }\textbf {\bibinfo {volume}
  {96}},\ \bibinfo {pages} {060502} (\bibinfo {year} {2006})}\BibitemShut
  {NoStop}%
\bibitem [{\citenamefont {Pirandola}\ \emph {et~al.}(2015)\citenamefont
  {Pirandola}, \citenamefont {Eisert}, \citenamefont {Weedbrook}, \citenamefont
  {Furusawa},\ and\ \citenamefont {Braunstein}}]{pirandola2015advances}%
  \BibitemOpen
  \bibfield  {author} {\bibinfo {author} {\bibfnamefont {S.}~\bibnamefont
  {Pirandola}}, \bibinfo {author} {\bibfnamefont {J.}~\bibnamefont {Eisert}},
  \bibinfo {author} {\bibfnamefont {C.}~\bibnamefont {Weedbrook}}, \bibinfo
  {author} {\bibfnamefont {A.}~\bibnamefont {Furusawa}},\ and\ \bibinfo
  {author} {\bibfnamefont {S.~L.}\ \bibnamefont {Braunstein}},\ }\bibfield
  {title} {\bibinfo {title} {Advances in quantum teleportation},\ }\href@noop
  {} {\bibfield  {journal} {\bibinfo  {journal} {Nature photonics}\ }\textbf
  {\bibinfo {volume} {9}},\ \bibinfo {pages} {641} (\bibinfo {year}
  {2015})}\BibitemShut {NoStop}%
\bibitem [{\citenamefont {Muralidharan}\ and\ \citenamefont
  {Panigrahi}(2008)}]{muralidharan2008perfect}%
  \BibitemOpen
  \bibfield  {author} {\bibinfo {author} {\bibfnamefont {S.}~\bibnamefont
  {Muralidharan}}\ and\ \bibinfo {author} {\bibfnamefont {P.~K.}\ \bibnamefont
  {Panigrahi}},\ }\bibfield  {title} {\bibinfo {title} {Perfect teleportation,
  quantum-state sharing, and superdense coding through a genuinely entangled
  five-qubit state},\ }\href@noop {} {\bibfield  {journal} {\bibinfo  {journal}
  {Physical Review A}\ }\textbf {\bibinfo {volume} {77}},\ \bibinfo {pages}
  {032321} (\bibinfo {year} {2008})}\BibitemShut {NoStop}%
\bibitem [{\citenamefont {Choudhury}\ \emph {et~al.}(2009)\citenamefont
  {Choudhury}, \citenamefont {Muralidharan},\ and\ \citenamefont
  {Panigrahi}}]{choudhury2009quantum}%
  \BibitemOpen
  \bibfield  {author} {\bibinfo {author} {\bibfnamefont {S.}~\bibnamefont
  {Choudhury}}, \bibinfo {author} {\bibfnamefont {S.}~\bibnamefont
  {Muralidharan}},\ and\ \bibinfo {author} {\bibfnamefont {P.~K.}\ \bibnamefont
  {Panigrahi}},\ }\bibfield  {title} {\bibinfo {title} {Quantum teleportation
  and state sharing using a genuinely entangled six-qubit state},\ }\href@noop
  {} {\bibfield  {journal} {\bibinfo  {journal} {Journal of Physics A:
  Mathematical and Theoretical}\ }\textbf {\bibinfo {volume} {42}},\ \bibinfo
  {pages} {115303} (\bibinfo {year} {2009})}\BibitemShut {NoStop}%
\bibitem [{\citenamefont {Saha}\ and\ \citenamefont
  {Panigrahi}(2012)}]{saha2012n}%
  \BibitemOpen
  \bibfield  {author} {\bibinfo {author} {\bibfnamefont {D.}~\bibnamefont
  {Saha}}\ and\ \bibinfo {author} {\bibfnamefont {P.~K.}\ \bibnamefont
  {Panigrahi}},\ }\bibfield  {title} {\bibinfo {title} {N-qubit quantum
  teleportation, information splitting and superdense coding through the
  composite ghz--bell channel},\ }\href@noop {} {\bibfield  {journal} {\bibinfo
   {journal} {Quantum Information Processing}\ }\textbf {\bibinfo {volume}
  {11}},\ \bibinfo {pages} {615} (\bibinfo {year} {2012})}\BibitemShut
  {NoStop}%
\bibitem [{\citenamefont {Bennett}\ \emph {et~al.}(1993)\citenamefont
  {Bennett}, \citenamefont {Brassard}, \citenamefont {Cr{\'e}peau},
  \citenamefont {Jozsa}, \citenamefont {Peres},\ and\ \citenamefont
  {Wootters}}]{bennett1993teleporting}%
  \BibitemOpen
  \bibfield  {author} {\bibinfo {author} {\bibfnamefont {C.~H.}\ \bibnamefont
  {Bennett}}, \bibinfo {author} {\bibfnamefont {G.}~\bibnamefont {Brassard}},
  \bibinfo {author} {\bibfnamefont {C.}~\bibnamefont {Cr{\'e}peau}}, \bibinfo
  {author} {\bibfnamefont {R.}~\bibnamefont {Jozsa}}, \bibinfo {author}
  {\bibfnamefont {A.}~\bibnamefont {Peres}},\ and\ \bibinfo {author}
  {\bibfnamefont {W.~K.}\ \bibnamefont {Wootters}},\ }\bibfield  {title}
  {\bibinfo {title} {Teleporting an unknown quantum state via dual classical
  and einstein-podolsky-rosen channels},\ }\href@noop {} {\bibfield  {journal}
  {\bibinfo  {journal} {Physical review letters}\ }\textbf {\bibinfo {volume}
  {70}},\ \bibinfo {pages} {1895} (\bibinfo {year} {1993})}\BibitemShut
  {NoStop}%
\bibitem [{\citenamefont {Karlsson}\ and\ \citenamefont
  {Bourennane}(1998)}]{karlsson1998quantum}%
  \BibitemOpen
  \bibfield  {author} {\bibinfo {author} {\bibfnamefont {A.}~\bibnamefont
  {Karlsson}}\ and\ \bibinfo {author} {\bibfnamefont {M.}~\bibnamefont
  {Bourennane}},\ }\bibfield  {title} {\bibinfo {title} {Quantum teleportation
  using three-particle entanglement},\ }\href@noop {} {\bibfield  {journal}
  {\bibinfo  {journal} {Physical Review A}\ }\textbf {\bibinfo {volume} {58}},\
  \bibinfo {pages} {4394} (\bibinfo {year} {1998})}\BibitemShut {NoStop}%
\bibitem [{\citenamefont {Kumar}\ \emph {et~al.}(2020)\citenamefont {Kumar},
  \citenamefont {Haddadi}, \citenamefont {Pourkarimi}, \citenamefont {Behera},\
  and\ \citenamefont {Panigrahi}}]{kumar2020experimental}%
  \BibitemOpen
  \bibfield  {author} {\bibinfo {author} {\bibfnamefont {A.}~\bibnamefont
  {Kumar}}, \bibinfo {author} {\bibfnamefont {S.}~\bibnamefont {Haddadi}},
  \bibinfo {author} {\bibfnamefont {M.~R.}\ \bibnamefont {Pourkarimi}},
  \bibinfo {author} {\bibfnamefont {B.~K.}\ \bibnamefont {Behera}},\ and\
  \bibinfo {author} {\bibfnamefont {P.~K.}\ \bibnamefont {Panigrahi}},\
  }\bibfield  {title} {\bibinfo {title} {Experimental realization of controlled
  quantum teleportation of arbitrary qubit states via cluster states},\
  }\href@noop {} {\bibfield  {journal} {\bibinfo  {journal} {Scientific
  Reports}\ }\textbf {\bibinfo {volume} {10}},\ \bibinfo {pages} {1} (\bibinfo
  {year} {2020})}\BibitemShut {NoStop}%
\bibitem [{\citenamefont {Agrawal}\ and\ \citenamefont
  {Pati}(2006)}]{agrawal2006perfect}%
  \BibitemOpen
  \bibfield  {author} {\bibinfo {author} {\bibfnamefont {P.}~\bibnamefont
  {Agrawal}}\ and\ \bibinfo {author} {\bibfnamefont {A.}~\bibnamefont {Pati}},\
  }\bibfield  {title} {\bibinfo {title} {Perfect teleportation and superdense
  coding with w states},\ }\href@noop {} {\bibfield  {journal} {\bibinfo
  {journal} {Physical Review A}\ }\textbf {\bibinfo {volume} {74}},\ \bibinfo
  {pages} {062320} (\bibinfo {year} {2006})}\BibitemShut {NoStop}%
\bibitem [{\citenamefont {Rao}\ \emph {et~al.}(2008)\citenamefont {Rao},
  \citenamefont {Ghosh},\ and\ \citenamefont {Panigrahi}}]{rao2008generation}%
  \BibitemOpen
  \bibfield  {author} {\bibinfo {author} {\bibfnamefont {D.~B.}\ \bibnamefont
  {Rao}}, \bibinfo {author} {\bibfnamefont {S.}~\bibnamefont {Ghosh}},\ and\
  \bibinfo {author} {\bibfnamefont {P.~K.}\ \bibnamefont {Panigrahi}},\
  }\bibfield  {title} {\bibinfo {title} {Generation of entangled channels for
  perfect teleportation using multielectron quantum dots},\ }\href@noop {}
  {\bibfield  {journal} {\bibinfo  {journal} {Physical Review A}\ }\textbf
  {\bibinfo {volume} {78}},\ \bibinfo {pages} {042328} (\bibinfo {year}
  {2008})}\BibitemShut {NoStop}%
\bibitem [{\citenamefont {Li}\ and\ \citenamefont {Qiu}(2007)}]{li2007states}%
  \BibitemOpen
  \bibfield  {author} {\bibinfo {author} {\bibfnamefont {L.}~\bibnamefont
  {Li}}\ and\ \bibinfo {author} {\bibfnamefont {D.}~\bibnamefont {Qiu}},\
  }\bibfield  {title} {\bibinfo {title} {The states of w-class as shared
  resources for perfect teleportation and superdense coding},\ }\href@noop {}
  {\bibfield  {journal} {\bibinfo  {journal} {Journal of Physics A:
  Mathematical and Theoretical}\ }\textbf {\bibinfo {volume} {40}},\ \bibinfo
  {pages} {10871} (\bibinfo {year} {2007})}\BibitemShut {NoStop}%
\bibitem [{\citenamefont {Li}\ \emph {et~al.}(2016)\citenamefont {Li},
  \citenamefont {Kong}, \citenamefont {Yang}, \citenamefont {Ozaydin},
  \citenamefont {Yang},\ and\ \citenamefont {Cao}}]{li2016generating}%
  \BibitemOpen
  \bibfield  {author} {\bibinfo {author} {\bibfnamefont {K.}~\bibnamefont
  {Li}}, \bibinfo {author} {\bibfnamefont {F.-Z.}\ \bibnamefont {Kong}},
  \bibinfo {author} {\bibfnamefont {M.}~\bibnamefont {Yang}}, \bibinfo {author}
  {\bibfnamefont {F.}~\bibnamefont {Ozaydin}}, \bibinfo {author} {\bibfnamefont
  {Q.}~\bibnamefont {Yang}},\ and\ \bibinfo {author} {\bibfnamefont {Z.-L.}\
  \bibnamefont {Cao}},\ }\bibfield  {title} {\bibinfo {title} {Generating
  multi-photon w-like states for perfect quantum teleportation and superdense
  coding},\ }\href@noop {} {\bibfield  {journal} {\bibinfo  {journal} {Quantum
  Information Processing}\ }\textbf {\bibinfo {volume} {15}},\ \bibinfo {pages}
  {3137} (\bibinfo {year} {2016})}\BibitemShut {NoStop}%
\bibitem [{\citenamefont {Zhao-Hui}\ \emph {et~al.}(2008)\citenamefont
  {Zhao-Hui}, \citenamefont {Chun-Xia},\ and\ \citenamefont
  {Jun-Gang}}]{zhao2008scheme}%
  \BibitemOpen
  \bibfield  {author} {\bibinfo {author} {\bibfnamefont {P.}~\bibnamefont
  {Zhao-Hui}}, \bibinfo {author} {\bibfnamefont {J.}~\bibnamefont {Chun-Xia}},\
  and\ \bibinfo {author} {\bibfnamefont {L.}~\bibnamefont {Jun-Gang}},\
  }\bibfield  {title} {\bibinfo {title} {Scheme for implementing perfect
  quantum teleportation with non-maximally entangled w-class state in cavity
  qed},\ }\href@noop {} {\bibfield  {journal} {\bibinfo  {journal}
  {Communications in Theoretical Physics}\ }\textbf {\bibinfo {volume} {50}},\
  \bibinfo {pages} {375} (\bibinfo {year} {2008})}\BibitemShut {NoStop}%
\bibitem [{\citenamefont {Swain}\ \emph
  {et~al.}(2020{\natexlab{a}})\citenamefont {Swain}, \citenamefont {Devrari},
  \citenamefont {Rai}, \citenamefont {Behera},\ and\ \citenamefont
  {Panigrahi}}]{swain2020generation}%
  \BibitemOpen
  \bibfield  {author} {\bibinfo {author} {\bibfnamefont {M.}~\bibnamefont
  {Swain}}, \bibinfo {author} {\bibfnamefont {V.}~\bibnamefont {Devrari}},
  \bibinfo {author} {\bibfnamefont {A.}~\bibnamefont {Rai}}, \bibinfo {author}
  {\bibfnamefont {B.~K.}\ \bibnamefont {Behera}},\ and\ \bibinfo {author}
  {\bibfnamefont {P.~K.}\ \bibnamefont {Panigrahi}},\ }\bibfield  {title}
  {\bibinfo {title} {Generation of perfect w-state and demonstration of its
  application to quantum information splitting},\ }\href@noop {} {\bibfield
  {journal} {\bibinfo  {journal} {arXiv preprint arXiv:2006.01742}\ } (\bibinfo
  {year} {2020}{\natexlab{a}})}\BibitemShut {NoStop}%
\bibitem [{\citenamefont {Van~Enk}(2005)}]{van2005single}%
  \BibitemOpen
  \bibfield  {author} {\bibinfo {author} {\bibfnamefont {S.}~\bibnamefont
  {Van~Enk}},\ }\bibfield  {title} {\bibinfo {title} {Single-particle
  entanglement},\ }\href@noop {} {\bibfield  {journal} {\bibinfo  {journal}
  {Physical Review A}\ }\textbf {\bibinfo {volume} {72}},\ \bibinfo {pages}
  {064306} (\bibinfo {year} {2005})}\BibitemShut {NoStop}%
\bibitem [{\citenamefont {Morin}\ \emph {et~al.}(2013)\citenamefont {Morin},
  \citenamefont {Bancal}, \citenamefont {Ho}, \citenamefont {Sekatski},
  \citenamefont {D’Auria}, \citenamefont {Gisin}, \citenamefont {Laurat},\
  and\ \citenamefont {Sangouard}}]{morin2013witnessing}%
  \BibitemOpen
  \bibfield  {author} {\bibinfo {author} {\bibfnamefont {O.}~\bibnamefont
  {Morin}}, \bibinfo {author} {\bibfnamefont {J.-D.}\ \bibnamefont {Bancal}},
  \bibinfo {author} {\bibfnamefont {M.}~\bibnamefont {Ho}}, \bibinfo {author}
  {\bibfnamefont {P.}~\bibnamefont {Sekatski}}, \bibinfo {author}
  {\bibfnamefont {V.}~\bibnamefont {D’Auria}}, \bibinfo {author}
  {\bibfnamefont {N.}~\bibnamefont {Gisin}}, \bibinfo {author} {\bibfnamefont
  {J.}~\bibnamefont {Laurat}},\ and\ \bibinfo {author} {\bibfnamefont
  {N.}~\bibnamefont {Sangouard}},\ }\bibfield  {title} {\bibinfo {title}
  {Witnessing trustworthy single-photon entanglement with local homodyne
  measurements},\ }\href@noop {} {\bibfield  {journal} {\bibinfo  {journal}
  {Physical review letters}\ }\textbf {\bibinfo {volume} {110}},\ \bibinfo
  {pages} {130401} (\bibinfo {year} {2013})}\BibitemShut {NoStop}%
\bibitem [{\citenamefont {Shi}\ \emph {et~al.}(2013)\citenamefont {Shi},
  \citenamefont {Xu}, \citenamefont {Zhong}, \citenamefont {Gong},
  \citenamefont {Bai}, \citenamefont {Yu}, \citenamefont {Li}, \citenamefont
  {Jin},\ and\ \citenamefont {Zhu}}]{shi2013heralded}%
  \BibitemOpen
  \bibfield  {author} {\bibinfo {author} {\bibfnamefont {J.}~\bibnamefont
  {Shi}}, \bibinfo {author} {\bibfnamefont {P.}~\bibnamefont {Xu}}, \bibinfo
  {author} {\bibfnamefont {M.}~\bibnamefont {Zhong}}, \bibinfo {author}
  {\bibfnamefont {Y.}~\bibnamefont {Gong}}, \bibinfo {author} {\bibfnamefont
  {Y.}~\bibnamefont {Bai}}, \bibinfo {author} {\bibfnamefont {W.}~\bibnamefont
  {Yu}}, \bibinfo {author} {\bibfnamefont {Q.}~\bibnamefont {Li}}, \bibinfo
  {author} {\bibfnamefont {H.}~\bibnamefont {Jin}},\ and\ \bibinfo {author}
  {\bibfnamefont {S.}~\bibnamefont {Zhu}},\ }\bibfield  {title} {\bibinfo
  {title} {Heralded generation of multipartite entanglement for one photon by
  using a single two-dimensional nonlinear photonic crystal},\ }\href@noop {}
  {\bibfield  {journal} {\bibinfo  {journal} {Optics express}\ }\textbf
  {\bibinfo {volume} {21}},\ \bibinfo {pages} {7875} (\bibinfo {year}
  {2013})}\BibitemShut {NoStop}%
\bibitem [{\citenamefont {Gr{\"a}fe}\ \emph {et~al.}(2014)\citenamefont
  {Gr{\"a}fe}, \citenamefont {Heilmann}, \citenamefont {Perez-Leija},
  \citenamefont {Keil}, \citenamefont {Dreisow}, \citenamefont {Heinrich},
  \citenamefont {Moya-Cessa}, \citenamefont {Nolte}, \citenamefont
  {Christodoulides},\ and\ \citenamefont {Szameit}}]{grafe2014chip}%
  \BibitemOpen
  \bibfield  {author} {\bibinfo {author} {\bibfnamefont {M.}~\bibnamefont
  {Gr{\"a}fe}}, \bibinfo {author} {\bibfnamefont {R.}~\bibnamefont {Heilmann}},
  \bibinfo {author} {\bibfnamefont {A.}~\bibnamefont {Perez-Leija}}, \bibinfo
  {author} {\bibfnamefont {R.}~\bibnamefont {Keil}}, \bibinfo {author}
  {\bibfnamefont {F.}~\bibnamefont {Dreisow}}, \bibinfo {author} {\bibfnamefont
  {M.}~\bibnamefont {Heinrich}}, \bibinfo {author} {\bibfnamefont
  {H.}~\bibnamefont {Moya-Cessa}}, \bibinfo {author} {\bibfnamefont
  {S.}~\bibnamefont {Nolte}}, \bibinfo {author} {\bibfnamefont {D.~N.}\
  \bibnamefont {Christodoulides}},\ and\ \bibinfo {author} {\bibfnamefont
  {A.}~\bibnamefont {Szameit}},\ }\bibfield  {title} {\bibinfo {title} {On-chip
  generation of high-order single-photon w-states},\ }\href@noop {} {\bibfield
  {journal} {\bibinfo  {journal} {Nature Photonics}\ }\textbf {\bibinfo
  {volume} {8}},\ \bibinfo {pages} {791} (\bibinfo {year} {2014})}\BibitemShut
  {NoStop}%
\bibitem [{\citenamefont {Monteiro}\ \emph {et~al.}(2015)\citenamefont
  {Monteiro}, \citenamefont {Vivoli}, \citenamefont {Guerreiro}, \citenamefont
  {Martin}, \citenamefont {Bancal}, \citenamefont {Zbinden}, \citenamefont
  {Thew},\ and\ \citenamefont {Sangouard}}]{monteiro2015revealing}%
  \BibitemOpen
  \bibfield  {author} {\bibinfo {author} {\bibfnamefont {F.}~\bibnamefont
  {Monteiro}}, \bibinfo {author} {\bibfnamefont {V.~C.}\ \bibnamefont
  {Vivoli}}, \bibinfo {author} {\bibfnamefont {T.}~\bibnamefont {Guerreiro}},
  \bibinfo {author} {\bibfnamefont {A.}~\bibnamefont {Martin}}, \bibinfo
  {author} {\bibfnamefont {J.-D.}\ \bibnamefont {Bancal}}, \bibinfo {author}
  {\bibfnamefont {H.}~\bibnamefont {Zbinden}}, \bibinfo {author} {\bibfnamefont
  {R.~T.}\ \bibnamefont {Thew}},\ and\ \bibinfo {author} {\bibfnamefont
  {N.}~\bibnamefont {Sangouard}},\ }\bibfield  {title} {\bibinfo {title}
  {Revealing genuine optical-path entanglement},\ }\href@noop {} {\bibfield
  {journal} {\bibinfo  {journal} {Physical review letters}\ }\textbf {\bibinfo
  {volume} {114}},\ \bibinfo {pages} {170504} (\bibinfo {year}
  {2015})}\BibitemShut {NoStop}%
\bibitem [{\citenamefont {Caspar}\ \emph {et~al.}(2020)\citenamefont {Caspar},
  \citenamefont {Verbanis}, \citenamefont {Oudot}, \citenamefont {Maring},
  \citenamefont {Samara}, \citenamefont {Caloz}, \citenamefont {Perrenoud},
  \citenamefont {Sekatski}, \citenamefont {Martin}, \citenamefont {Sangouard}
  \emph {et~al.}}]{caspar2020heralded}%
  \BibitemOpen
  \bibfield  {author} {\bibinfo {author} {\bibfnamefont {P.}~\bibnamefont
  {Caspar}}, \bibinfo {author} {\bibfnamefont {E.}~\bibnamefont {Verbanis}},
  \bibinfo {author} {\bibfnamefont {E.}~\bibnamefont {Oudot}}, \bibinfo
  {author} {\bibfnamefont {N.}~\bibnamefont {Maring}}, \bibinfo {author}
  {\bibfnamefont {F.}~\bibnamefont {Samara}}, \bibinfo {author} {\bibfnamefont
  {M.}~\bibnamefont {Caloz}}, \bibinfo {author} {\bibfnamefont
  {M.}~\bibnamefont {Perrenoud}}, \bibinfo {author} {\bibfnamefont
  {P.}~\bibnamefont {Sekatski}}, \bibinfo {author} {\bibfnamefont
  {A.}~\bibnamefont {Martin}}, \bibinfo {author} {\bibfnamefont
  {N.}~\bibnamefont {Sangouard}}, \emph {et~al.},\ }\bibfield  {title}
  {\bibinfo {title} {Heralded distribution of single-photon path
  entanglement},\ }\href@noop {} {\bibfield  {journal} {\bibinfo  {journal}
  {Physical Review Letters}\ }\textbf {\bibinfo {volume} {125}},\ \bibinfo
  {pages} {110506} (\bibinfo {year} {2020})}\BibitemShut {NoStop}%
\bibitem [{\citenamefont {Perez-Leija}\ \emph {et~al.}(2013)\citenamefont
  {Perez-Leija}, \citenamefont {Hernandez-Herrejon}, \citenamefont
  {Moya-Cessa}, \citenamefont {Szameit},\ and\ \citenamefont
  {Christodoulides}}]{perez2013generating}%
  \BibitemOpen
  \bibfield  {author} {\bibinfo {author} {\bibfnamefont {A.}~\bibnamefont
  {Perez-Leija}}, \bibinfo {author} {\bibfnamefont {J.}~\bibnamefont
  {Hernandez-Herrejon}}, \bibinfo {author} {\bibfnamefont {H.}~\bibnamefont
  {Moya-Cessa}}, \bibinfo {author} {\bibfnamefont {A.}~\bibnamefont
  {Szameit}},\ and\ \bibinfo {author} {\bibfnamefont {D.~N.}\ \bibnamefont
  {Christodoulides}},\ }\bibfield  {title} {\bibinfo {title} {Generating
  photon-encoded w states in multiport waveguide-array systems},\ }\href@noop
  {} {\bibfield  {journal} {\bibinfo  {journal} {Physical Review A}\ }\textbf
  {\bibinfo {volume} {87}},\ \bibinfo {pages} {013842} (\bibinfo {year}
  {2013})}\BibitemShut {NoStop}%
\bibitem [{\citenamefont {Hessmo}\ \emph {et~al.}(2004)\citenamefont {Hessmo},
  \citenamefont {Usachev}, \citenamefont {Heydari},\ and\ \citenamefont
  {Bj{\"o}rk}}]{hessmo2004experimental}%
  \BibitemOpen
  \bibfield  {author} {\bibinfo {author} {\bibfnamefont {B.}~\bibnamefont
  {Hessmo}}, \bibinfo {author} {\bibfnamefont {P.}~\bibnamefont {Usachev}},
  \bibinfo {author} {\bibfnamefont {H.}~\bibnamefont {Heydari}},\ and\ \bibinfo
  {author} {\bibfnamefont {G.}~\bibnamefont {Bj{\"o}rk}},\ }\bibfield  {title}
  {\bibinfo {title} {Experimental demonstration of single photon nonlocality},\
  }\href@noop {} {\bibfield  {journal} {\bibinfo  {journal} {Physical review
  letters}\ }\textbf {\bibinfo {volume} {92}},\ \bibinfo {pages} {180401}
  (\bibinfo {year} {2004})}\BibitemShut {NoStop}%
\bibitem [{\citenamefont {White}\ \emph {et~al.}(2020)\citenamefont {White},
  \citenamefont {Klauck}, \citenamefont {Tran}, \citenamefont {Schmitt},
  \citenamefont {Kianinia}, \citenamefont {Steinfurth}, \citenamefont
  {Heinrich}, \citenamefont {Toth}, \citenamefont {Szameit}, \citenamefont
  {Aharonovich} \emph {et~al.}}]{white2020quantum}%
  \BibitemOpen
  \bibfield  {author} {\bibinfo {author} {\bibfnamefont {S.~J.}\ \bibnamefont
  {White}}, \bibinfo {author} {\bibfnamefont {F.}~\bibnamefont {Klauck}},
  \bibinfo {author} {\bibfnamefont {T.~T.}\ \bibnamefont {Tran}}, \bibinfo
  {author} {\bibfnamefont {N.}~\bibnamefont {Schmitt}}, \bibinfo {author}
  {\bibfnamefont {M.}~\bibnamefont {Kianinia}}, \bibinfo {author}
  {\bibfnamefont {A.}~\bibnamefont {Steinfurth}}, \bibinfo {author}
  {\bibfnamefont {M.}~\bibnamefont {Heinrich}}, \bibinfo {author}
  {\bibfnamefont {M.}~\bibnamefont {Toth}}, \bibinfo {author} {\bibfnamefont
  {A.}~\bibnamefont {Szameit}}, \bibinfo {author} {\bibfnamefont
  {I.}~\bibnamefont {Aharonovich}}, \emph {et~al.},\ }\bibfield  {title}
  {\bibinfo {title} {Quantum random number generation using a hexagonal boron
  nitride single photon emitter},\ }\href@noop {} {\bibfield  {journal}
  {\bibinfo  {journal} {Journal of Optics}\ }\textbf {\bibinfo {volume} {23}},\
  \bibinfo {pages} {01LT01} (\bibinfo {year} {2020})}\BibitemShut {NoStop}%
\bibitem [{\citenamefont {Chen}\ \emph {et~al.}(2019)\citenamefont {Chen},
  \citenamefont {Greiner}, \citenamefont {Wrachtrup},\ and\ \citenamefont
  {Gerhardt}}]{chen2019single}%
  \BibitemOpen
  \bibfield  {author} {\bibinfo {author} {\bibfnamefont {X.}~\bibnamefont
  {Chen}}, \bibinfo {author} {\bibfnamefont {J.~N.}\ \bibnamefont {Greiner}},
  \bibinfo {author} {\bibfnamefont {J.}~\bibnamefont {Wrachtrup}},\ and\
  \bibinfo {author} {\bibfnamefont {I.}~\bibnamefont {Gerhardt}},\ }\bibfield
  {title} {\bibinfo {title} {Single photon randomness based on a defect center
  in diamond},\ }\href@noop {} {\bibfield  {journal} {\bibinfo  {journal}
  {Scientific reports}\ }\textbf {\bibinfo {volume} {9}},\ \bibinfo {pages} {1}
  (\bibinfo {year} {2019})}\BibitemShut {NoStop}%
\bibitem [{\citenamefont {Luo}\ \emph {et~al.}(2020)\citenamefont {Luo},
  \citenamefont {Cheng}, \citenamefont {Fan}, \citenamefont {Tan},
  \citenamefont {Song}, \citenamefont {Deng}, \citenamefont {Wang},\ and\
  \citenamefont {Zhou}}]{luo2020quantum}%
  \BibitemOpen
  \bibfield  {author} {\bibinfo {author} {\bibfnamefont {Q.}~\bibnamefont
  {Luo}}, \bibinfo {author} {\bibfnamefont {Z.}~\bibnamefont {Cheng}}, \bibinfo
  {author} {\bibfnamefont {J.}~\bibnamefont {Fan}}, \bibinfo {author}
  {\bibfnamefont {L.}~\bibnamefont {Tan}}, \bibinfo {author} {\bibfnamefont
  {H.}~\bibnamefont {Song}}, \bibinfo {author} {\bibfnamefont {G.}~\bibnamefont
  {Deng}}, \bibinfo {author} {\bibfnamefont {Y.}~\bibnamefont {Wang}},\ and\
  \bibinfo {author} {\bibfnamefont {Q.}~\bibnamefont {Zhou}},\ }\bibfield
  {title} {\bibinfo {title} {Quantum random number generator based on
  single-photon emitter in gallium nitride},\ }\href@noop {} {\bibfield
  {journal} {\bibinfo  {journal} {Optics Letters}\ }\textbf {\bibinfo {volume}
  {45}},\ \bibinfo {pages} {4224} (\bibinfo {year} {2020})}\BibitemShut
  {NoStop}%
\bibitem [{\citenamefont {Gottesman}\ \emph {et~al.}(2012)\citenamefont
  {Gottesman}, \citenamefont {Jennewein},\ and\ \citenamefont
  {Croke}}]{gottesman2012longer}%
  \BibitemOpen
  \bibfield  {author} {\bibinfo {author} {\bibfnamefont {D.}~\bibnamefont
  {Gottesman}}, \bibinfo {author} {\bibfnamefont {T.}~\bibnamefont
  {Jennewein}},\ and\ \bibinfo {author} {\bibfnamefont {S.}~\bibnamefont
  {Croke}},\ }\bibfield  {title} {\bibinfo {title} {Longer-baseline telescopes
  using quantum repeaters},\ }\href@noop {} {\bibfield  {journal} {\bibinfo
  {journal} {Physical review letters}\ }\textbf {\bibinfo {volume} {109}},\
  \bibinfo {pages} {070503} (\bibinfo {year} {2012})}\BibitemShut {NoStop}%
\bibitem [{\citenamefont {Rai}\ and\ \citenamefont
  {Agarwal}(2009)}]{rai2009possibility}%
  \BibitemOpen
  \bibfield  {author} {\bibinfo {author} {\bibfnamefont {A.}~\bibnamefont
  {Rai}}\ and\ \bibinfo {author} {\bibfnamefont {G.}~\bibnamefont {Agarwal}},\
  }\bibfield  {title} {\bibinfo {title} {Possibility of coherent phenomena such
  as bloch oscillations with single photons via w states},\ }\href@noop {}
  {\bibfield  {journal} {\bibinfo  {journal} {Physical Review A}\ }\textbf
  {\bibinfo {volume} {79}},\ \bibinfo {pages} {053849} (\bibinfo {year}
  {2009})}\BibitemShut {NoStop}%
\bibitem [{\citenamefont {Szameit}\ and\ \citenamefont
  {Nolte}(2010)}]{szameit2010discrete}%
  \BibitemOpen
  \bibfield  {author} {\bibinfo {author} {\bibfnamefont {A.}~\bibnamefont
  {Szameit}}\ and\ \bibinfo {author} {\bibfnamefont {S.}~\bibnamefont
  {Nolte}},\ }\bibfield  {title} {\bibinfo {title} {Discrete optics in
  femtosecond-laser-written photonic structures},\ }\href@noop {} {\bibfield
  {journal} {\bibinfo  {journal} {Journal of Physics B: Atomic, Molecular and
  Optical Physics}\ }\textbf {\bibinfo {volume} {43}},\ \bibinfo {pages}
  {163001} (\bibinfo {year} {2010})}\BibitemShut {NoStop}%
\bibitem [{\citenamefont {Meany}\ \emph {et~al.}(2015)\citenamefont {Meany},
  \citenamefont {Gr{\"a}fe}, \citenamefont {Heilmann}, \citenamefont
  {Perez-Leija}, \citenamefont {Gross}, \citenamefont {Steel}, \citenamefont
  {Withford},\ and\ \citenamefont {Szameit}}]{meany2015laser}%
  \BibitemOpen
  \bibfield  {author} {\bibinfo {author} {\bibfnamefont {T.}~\bibnamefont
  {Meany}}, \bibinfo {author} {\bibfnamefont {M.}~\bibnamefont {Gr{\"a}fe}},
  \bibinfo {author} {\bibfnamefont {R.}~\bibnamefont {Heilmann}}, \bibinfo
  {author} {\bibfnamefont {A.}~\bibnamefont {Perez-Leija}}, \bibinfo {author}
  {\bibfnamefont {S.}~\bibnamefont {Gross}}, \bibinfo {author} {\bibfnamefont
  {M.~J.}\ \bibnamefont {Steel}}, \bibinfo {author} {\bibfnamefont {M.~J.}\
  \bibnamefont {Withford}},\ and\ \bibinfo {author} {\bibfnamefont
  {A.}~\bibnamefont {Szameit}},\ }\bibfield  {title} {\bibinfo {title} {Laser
  written circuits for quantum photonics},\ }\href@noop {} {\bibfield
  {journal} {\bibinfo  {journal} {Laser \& Photonics Reviews}\ }\textbf
  {\bibinfo {volume} {9}},\ \bibinfo {pages} {363} (\bibinfo {year}
  {2015})}\BibitemShut {NoStop}%
\bibitem [{\citenamefont {Keil}\ \emph {et~al.}(2015)\citenamefont {Keil},
  \citenamefont {Noh}, \citenamefont {Rai}, \citenamefont {St{\"u}tzer},
  \citenamefont {Nolte}, \citenamefont {Angelakis},\ and\ \citenamefont
  {Szameit}}]{keil2015optical}%
  \BibitemOpen
  \bibfield  {author} {\bibinfo {author} {\bibfnamefont {R.}~\bibnamefont
  {Keil}}, \bibinfo {author} {\bibfnamefont {C.}~\bibnamefont {Noh}}, \bibinfo
  {author} {\bibfnamefont {A.}~\bibnamefont {Rai}}, \bibinfo {author}
  {\bibfnamefont {S.}~\bibnamefont {St{\"u}tzer}}, \bibinfo {author}
  {\bibfnamefont {S.}~\bibnamefont {Nolte}}, \bibinfo {author} {\bibfnamefont
  {D.~G.}\ \bibnamefont {Angelakis}},\ and\ \bibinfo {author} {\bibfnamefont
  {A.}~\bibnamefont {Szameit}},\ }\bibfield  {title} {\bibinfo {title} {Optical
  simulation of charge conservation violation and majorana dynamics},\
  }\href@noop {} {\bibfield  {journal} {\bibinfo  {journal} {Optica}\ }\textbf
  {\bibinfo {volume} {2}},\ \bibinfo {pages} {454} (\bibinfo {year}
  {2015})}\BibitemShut {NoStop}%
\bibitem [{\citenamefont {Rai}\ \emph {et~al.}(2015)\citenamefont {Rai},
  \citenamefont {Lee}, \citenamefont {Noh},\ and\ \citenamefont
  {Angelakis}}]{rai2015photonic}%
  \BibitemOpen
  \bibfield  {author} {\bibinfo {author} {\bibfnamefont {A.}~\bibnamefont
  {Rai}}, \bibinfo {author} {\bibfnamefont {C.}~\bibnamefont {Lee}}, \bibinfo
  {author} {\bibfnamefont {C.}~\bibnamefont {Noh}},\ and\ \bibinfo {author}
  {\bibfnamefont {D.~G.}\ \bibnamefont {Angelakis}},\ }\bibfield  {title}
  {\bibinfo {title} {Photonic lattice simulation of dissipation-induced
  correlations in bosonic systems},\ }\href@noop {} {\bibfield  {journal}
  {\bibinfo  {journal} {Scientific Reports}\ }\textbf {\bibinfo {volume} {5}},\
  \bibinfo {pages} {1} (\bibinfo {year} {2015})}\BibitemShut {NoStop}%
\bibitem [{\citenamefont {Heilmann}\ \emph {et~al.}(2015)\citenamefont
  {Heilmann}, \citenamefont {Gr{\"a}fe}, \citenamefont {Nolte},\ and\
  \citenamefont {Szameit}}]{heilmann2015novel}%
  \BibitemOpen
  \bibfield  {author} {\bibinfo {author} {\bibfnamefont {R.}~\bibnamefont
  {Heilmann}}, \bibinfo {author} {\bibfnamefont {M.}~\bibnamefont {Gr{\"a}fe}},
  \bibinfo {author} {\bibfnamefont {S.}~\bibnamefont {Nolte}},\ and\ \bibinfo
  {author} {\bibfnamefont {A.}~\bibnamefont {Szameit}},\ }\bibfield  {title}
  {\bibinfo {title} {A novel integrated quantum circuit for high-order w-state
  generation and its highly precise characterization},\ }\href@noop {}
  {\bibfield  {journal} {\bibinfo  {journal} {Science bulletin}\ }\textbf
  {\bibinfo {volume} {60}},\ \bibinfo {pages} {96} (\bibinfo {year}
  {2015})}\BibitemShut {NoStop}%
\bibitem [{\citenamefont {Rai}\ \emph {et~al.}(2010)\citenamefont {Rai},
  \citenamefont {Das},\ and\ \citenamefont {Agarwal}}]{rai2010quantum}%
  \BibitemOpen
  \bibfield  {author} {\bibinfo {author} {\bibfnamefont {A.}~\bibnamefont
  {Rai}}, \bibinfo {author} {\bibfnamefont {S.}~\bibnamefont {Das}},\ and\
  \bibinfo {author} {\bibfnamefont {G.}~\bibnamefont {Agarwal}},\ }\bibfield
  {title} {\bibinfo {title} {Quantum entanglement in coupled lossy
  waveguides},\ }\href@noop {} {\bibfield  {journal} {\bibinfo  {journal}
  {Optics express}\ }\textbf {\bibinfo {volume} {18}},\ \bibinfo {pages} {6241}
  (\bibinfo {year} {2010})}\BibitemShut {NoStop}%
\bibitem [{\citenamefont {Swain}\ \emph
  {et~al.}(2020{\natexlab{b}})\citenamefont {Swain}, \citenamefont {Rai},
  \citenamefont {Selvan},\ and\ \citenamefont {Panigrahi}}]{swain2020single}%
  \BibitemOpen
  \bibfield  {author} {\bibinfo {author} {\bibfnamefont {M.}~\bibnamefont
  {Swain}}, \bibinfo {author} {\bibfnamefont {A.}~\bibnamefont {Rai}}, \bibinfo
  {author} {\bibfnamefont {M.~K.}\ \bibnamefont {Selvan}},\ and\ \bibinfo
  {author} {\bibfnamefont {P.~K.}\ \bibnamefont {Panigrahi}},\ }\bibfield
  {title} {\bibinfo {title} {Single photon generation and non-locality of
  perfect w-state},\ }\href@noop {} {\bibfield  {journal} {\bibinfo  {journal}
  {Journal of Optics}\ }\textbf {\bibinfo {volume} {22}},\ \bibinfo {pages}
  {075202} (\bibinfo {year} {2020}{\natexlab{b}})}\BibitemShut {NoStop}%
\bibitem [{\citenamefont {Wootters}(1998)}]{wootters1998entanglement}%
  \BibitemOpen
  \bibfield  {author} {\bibinfo {author} {\bibfnamefont {W.~K.}\ \bibnamefont
  {Wootters}},\ }\bibfield  {title} {\bibinfo {title} {Entanglement of
  formation of an arbitrary state of two qubits},\ }\href@noop {} {\bibfield
  {journal} {\bibinfo  {journal} {Physical Review Letters}\ }\textbf {\bibinfo
  {volume} {80}},\ \bibinfo {pages} {2245} (\bibinfo {year}
  {1998})}\BibitemShut {NoStop}%
\bibitem [{\citenamefont {Horodecki}\ \emph {et~al.}(2009)\citenamefont
  {Horodecki}, \citenamefont {Horodecki}, \citenamefont {Horodecki},\ and\
  \citenamefont {Horodecki}}]{horodecki2009quantum}%
  \BibitemOpen
  \bibfield  {author} {\bibinfo {author} {\bibfnamefont {R.}~\bibnamefont
  {Horodecki}}, \bibinfo {author} {\bibfnamefont {P.}~\bibnamefont
  {Horodecki}}, \bibinfo {author} {\bibfnamefont {M.}~\bibnamefont
  {Horodecki}},\ and\ \bibinfo {author} {\bibfnamefont {K.}~\bibnamefont
  {Horodecki}},\ }\bibfield  {title} {\bibinfo {title} {Quantum entanglement},\
  }\href@noop {} {\bibfield  {journal} {\bibinfo  {journal} {Reviews of modern
  physics}\ }\textbf {\bibinfo {volume} {81}},\ \bibinfo {pages} {865}
  (\bibinfo {year} {2009})}\BibitemShut {NoStop}%
\bibitem [{\citenamefont {Ma}\ \emph {et~al.}(2011)\citenamefont {Ma},
  \citenamefont {Chen}, \citenamefont {Chen}, \citenamefont {Spengler},
  \citenamefont {Gabriel},\ and\ \citenamefont {Huber}}]{ma2011measure}%
  \BibitemOpen
  \bibfield  {author} {\bibinfo {author} {\bibfnamefont {Z.-H.}\ \bibnamefont
  {Ma}}, \bibinfo {author} {\bibfnamefont {Z.-H.}\ \bibnamefont {Chen}},
  \bibinfo {author} {\bibfnamefont {J.-L.}\ \bibnamefont {Chen}}, \bibinfo
  {author} {\bibfnamefont {C.}~\bibnamefont {Spengler}}, \bibinfo {author}
  {\bibfnamefont {A.}~\bibnamefont {Gabriel}},\ and\ \bibinfo {author}
  {\bibfnamefont {M.}~\bibnamefont {Huber}},\ }\bibfield  {title} {\bibinfo
  {title} {Measure of genuine multipartite entanglement with computable lower
  bounds},\ }\href@noop {} {\bibfield  {journal} {\bibinfo  {journal} {Physical
  Review A}\ }\textbf {\bibinfo {volume} {83}},\ \bibinfo {pages} {062325}
  (\bibinfo {year} {2011})}\BibitemShut {NoStop}%
\bibitem [{\citenamefont {Rafsanjani}\ \emph {et~al.}(2012)\citenamefont
  {Rafsanjani}, \citenamefont {Huber}, \citenamefont {Broadbent},\ and\
  \citenamefont {Eberly}}]{rafsanjani2012genuinely}%
  \BibitemOpen
  \bibfield  {author} {\bibinfo {author} {\bibfnamefont {S.~H.}\ \bibnamefont
  {Rafsanjani}}, \bibinfo {author} {\bibfnamefont {M.}~\bibnamefont {Huber}},
  \bibinfo {author} {\bibfnamefont {C.~J.}\ \bibnamefont {Broadbent}},\ and\
  \bibinfo {author} {\bibfnamefont {J.~H.}\ \bibnamefont {Eberly}},\ }\bibfield
   {title} {\bibinfo {title} {Genuinely multipartite concurrence of n-qubit x
  matrices},\ }\href@noop {} {\bibfield  {journal} {\bibinfo  {journal}
  {Physical Review A}\ }\textbf {\bibinfo {volume} {86}},\ \bibinfo {pages}
  {062303} (\bibinfo {year} {2012})}\BibitemShut {NoStop}%
\bibitem [{\citenamefont {Coffman}\ \emph {et~al.}(2000)\citenamefont
  {Coffman}, \citenamefont {Kundu},\ and\ \citenamefont
  {Wootters}}]{coffman2000distributed}%
  \BibitemOpen
  \bibfield  {author} {\bibinfo {author} {\bibfnamefont {V.}~\bibnamefont
  {Coffman}}, \bibinfo {author} {\bibfnamefont {J.}~\bibnamefont {Kundu}},\
  and\ \bibinfo {author} {\bibfnamefont {W.~K.}\ \bibnamefont {Wootters}},\
  }\bibfield  {title} {\bibinfo {title} {Distributed entanglement},\
  }\href@noop {} {\bibfield  {journal} {\bibinfo  {journal} {Physical Review
  A}\ }\textbf {\bibinfo {volume} {61}},\ \bibinfo {pages} {052306} (\bibinfo
  {year} {2000})}\BibitemShut {NoStop}%
\bibitem [{\citenamefont {Koashi}\ and\ \citenamefont
  {Winter}(2004)}]{koashi2004monogamy}%
  \BibitemOpen
  \bibfield  {author} {\bibinfo {author} {\bibfnamefont {M.}~\bibnamefont
  {Koashi}}\ and\ \bibinfo {author} {\bibfnamefont {A.}~\bibnamefont
  {Winter}},\ }\bibfield  {title} {\bibinfo {title} {Monogamy of quantum
  entanglement and other correlations},\ }\href@noop {} {\bibfield  {journal}
  {\bibinfo  {journal} {Physical Review A}\ }\textbf {\bibinfo {volume} {69}},\
  \bibinfo {pages} {022309} (\bibinfo {year} {2004})}\BibitemShut {NoStop}%
\bibitem [{\citenamefont {Zhu}\ and\ \citenamefont
  {Fei}(2015)}]{zhu2015generalized}%
  \BibitemOpen
  \bibfield  {author} {\bibinfo {author} {\bibfnamefont {X.-N.}\ \bibnamefont
  {Zhu}}\ and\ \bibinfo {author} {\bibfnamefont {S.-M.}\ \bibnamefont {Fei}},\
  }\bibfield  {title} {\bibinfo {title} {Generalized monogamy relations of
  concurrence for n-qubit systems},\ }\href@noop {} {\bibfield  {journal}
  {\bibinfo  {journal} {Physical Review A}\ }\textbf {\bibinfo {volume} {92}},\
  \bibinfo {pages} {062345} (\bibinfo {year} {2015})}\BibitemShut {NoStop}%
\bibitem [{\citenamefont {Peres}(1996)}]{peres1996separability}%
  \BibitemOpen
  \bibfield  {author} {\bibinfo {author} {\bibfnamefont {A.}~\bibnamefont
  {Peres}},\ }\bibfield  {title} {\bibinfo {title} {Separability criterion for
  density matrices},\ }\href@noop {} {\bibfield  {journal} {\bibinfo  {journal}
  {Physical Review Letters}\ }\textbf {\bibinfo {volume} {77}},\ \bibinfo
  {pages} {1413} (\bibinfo {year} {1996})}\BibitemShut {NoStop}%
\bibitem [{\citenamefont {Horodecki}\ \emph {et~al.}(1996)\citenamefont
  {Horodecki}, \citenamefont {Horodecki},\ and\ \citenamefont
  {Horodecki}}]{horodecki1996necessary}%
  \BibitemOpen
  \bibfield  {author} {\bibinfo {author} {\bibfnamefont {M.}~\bibnamefont
  {Horodecki}}, \bibinfo {author} {\bibfnamefont {P.}~\bibnamefont
  {Horodecki}},\ and\ \bibinfo {author} {\bibfnamefont {R.}~\bibnamefont
  {Horodecki}},\ }\bibfield  {title} {\bibinfo {title} {On the necessary and
  sufficient conditions for separability of mixed quantum states},\ }\href@noop
  {} {\bibfield  {journal} {\bibinfo  {journal} {Phys. Lett. A}\ }\textbf
  {\bibinfo {volume} {223}} (\bibinfo {year} {1996})}\BibitemShut {NoStop}%
\bibitem [{\citenamefont {Terhal}(2002)}]{terhal2002detecting}%
  \BibitemOpen
  \bibfield  {author} {\bibinfo {author} {\bibfnamefont {B.~M.}\ \bibnamefont
  {Terhal}},\ }\bibfield  {title} {\bibinfo {title} {Detecting quantum
  entanglement},\ }\href@noop {} {\bibfield  {journal} {\bibinfo  {journal}
  {Theoretical Computer Science}\ }\textbf {\bibinfo {volume} {287}},\ \bibinfo
  {pages} {313} (\bibinfo {year} {2002})}\BibitemShut {NoStop}%
\bibitem [{\citenamefont {Bourennane}\ \emph {et~al.}(2004)\citenamefont
  {Bourennane}, \citenamefont {Eibl}, \citenamefont {Kurtsiefer}, \citenamefont
  {Gaertner}, \citenamefont {Weinfurter}, \citenamefont {G{\"u}hne},
  \citenamefont {Hyllus}, \citenamefont {Bru{\ss}}, \citenamefont
  {Lewenstein},\ and\ \citenamefont {Sanpera}}]{bourennane2004experimental}%
  \BibitemOpen
  \bibfield  {author} {\bibinfo {author} {\bibfnamefont {M.}~\bibnamefont
  {Bourennane}}, \bibinfo {author} {\bibfnamefont {M.}~\bibnamefont {Eibl}},
  \bibinfo {author} {\bibfnamefont {C.}~\bibnamefont {Kurtsiefer}}, \bibinfo
  {author} {\bibfnamefont {S.}~\bibnamefont {Gaertner}}, \bibinfo {author}
  {\bibfnamefont {H.}~\bibnamefont {Weinfurter}}, \bibinfo {author}
  {\bibfnamefont {O.}~\bibnamefont {G{\"u}hne}}, \bibinfo {author}
  {\bibfnamefont {P.}~\bibnamefont {Hyllus}}, \bibinfo {author} {\bibfnamefont
  {D.}~\bibnamefont {Bru{\ss}}}, \bibinfo {author} {\bibfnamefont
  {M.}~\bibnamefont {Lewenstein}},\ and\ \bibinfo {author} {\bibfnamefont
  {A.}~\bibnamefont {Sanpera}},\ }\bibfield  {title} {\bibinfo {title}
  {Experimental detection of multipartite entanglement using witness
  operators},\ }\href@noop {} {\bibfield  {journal} {\bibinfo  {journal}
  {Physical review letters}\ }\textbf {\bibinfo {volume} {92}},\ \bibinfo
  {pages} {087902} (\bibinfo {year} {2004})}\BibitemShut {NoStop}%
\bibitem [{\citenamefont {G{\"u}hne}\ and\ \citenamefont
  {T{\'o}th}(2009)}]{guhne2009entanglement}%
  \BibitemOpen
  \bibfield  {author} {\bibinfo {author} {\bibfnamefont {O.}~\bibnamefont
  {G{\"u}hne}}\ and\ \bibinfo {author} {\bibfnamefont {G.}~\bibnamefont
  {T{\'o}th}},\ }\bibfield  {title} {\bibinfo {title} {Entanglement
  detection},\ }\href@noop {} {\bibfield  {journal} {\bibinfo  {journal}
  {Physics Reports}\ }\textbf {\bibinfo {volume} {474}},\ \bibinfo {pages} {1}
  (\bibinfo {year} {2009})}\BibitemShut {NoStop}%
\bibitem [{\citenamefont {Agarwal}\ and\ \citenamefont
  {Biswas}(2005)}]{agarwal2005inseparability}%
  \BibitemOpen
  \bibfield  {author} {\bibinfo {author} {\bibfnamefont {G.~S.}\ \bibnamefont
  {Agarwal}}\ and\ \bibinfo {author} {\bibfnamefont {A.}~\bibnamefont
  {Biswas}},\ }\bibfield  {title} {\bibinfo {title} {Inseparability
  inequalities for higher order moments for bipartite systems},\ }\href@noop {}
  {\bibfield  {journal} {\bibinfo  {journal} {New Journal of Physics}\ }\textbf
  {\bibinfo {volume} {7}},\ \bibinfo {pages} {211} (\bibinfo {year}
  {2005})}\BibitemShut {NoStop}%
\bibitem [{\citenamefont {Hillery}\ and\ \citenamefont
  {Zubairy}(2006{\natexlab{a}})}]{hillery2006entanglement}%
  \BibitemOpen
  \bibfield  {author} {\bibinfo {author} {\bibfnamefont {M.}~\bibnamefont
  {Hillery}}\ and\ \bibinfo {author} {\bibfnamefont {M.~S.}\ \bibnamefont
  {Zubairy}},\ }\bibfield  {title} {\bibinfo {title} {Entanglement conditions
  for two-mode states},\ }\href@noop {} {\bibfield  {journal} {\bibinfo
  {journal} {Physical review letters}\ }\textbf {\bibinfo {volume} {96}},\
  \bibinfo {pages} {050503} (\bibinfo {year} {2006}{\natexlab{a}})}\BibitemShut
  {NoStop}%
\bibitem [{\citenamefont {Hillery}\ and\ \citenamefont
  {Zubairy}(2006{\natexlab{b}})}]{hillery2006entanglement1}%
  \BibitemOpen
  \bibfield  {author} {\bibinfo {author} {\bibfnamefont {M.}~\bibnamefont
  {Hillery}}\ and\ \bibinfo {author} {\bibfnamefont {M.~S.}\ \bibnamefont
  {Zubairy}},\ }\bibfield  {title} {\bibinfo {title} {Entanglement conditions
  for two-mode states: applications},\ }\href@noop {} {\bibfield  {journal}
  {\bibinfo  {journal} {Physical Review A}\ }\textbf {\bibinfo {volume} {74}},\
  \bibinfo {pages} {032333} (\bibinfo {year} {2006}{\natexlab{b}})}\BibitemShut
  {NoStop}%
\bibitem [{\citenamefont {Nha}\ and\ \citenamefont
  {Kim}(2006)}]{nha2006entanglement}%
  \BibitemOpen
  \bibfield  {author} {\bibinfo {author} {\bibfnamefont {H.}~\bibnamefont
  {Nha}}\ and\ \bibinfo {author} {\bibfnamefont {J.}~\bibnamefont {Kim}},\
  }\bibfield  {title} {\bibinfo {title} {Entanglement criteria via the
  uncertainty relations in su (2) and su (1, 1) algebras: Detection of
  non-gaussian entangled states},\ }\href@noop {} {\bibfield  {journal}
  {\bibinfo  {journal} {Physical Review A}\ }\textbf {\bibinfo {volume} {74}},\
  \bibinfo {pages} {012317} (\bibinfo {year} {2006})}\BibitemShut {NoStop}%
\bibitem [{\citenamefont {Nha}(2008)}]{nha2008linear}%
  \BibitemOpen
  \bibfield  {author} {\bibinfo {author} {\bibfnamefont {H.}~\bibnamefont
  {Nha}},\ }\bibfield  {title} {\bibinfo {title} {Linear optical scheme to
  demonstrate genuine multipartite entanglement for single-particle w states},\
  }\href@noop {} {\bibfield  {journal} {\bibinfo  {journal} {Physical Review
  A}\ }\textbf {\bibinfo {volume} {77}},\ \bibinfo {pages} {062328} (\bibinfo
  {year} {2008})}\BibitemShut {NoStop}%
\bibitem [{\citenamefont {Selvan}\ and\ \citenamefont
  {Panigrahi}(2019)}]{selvan2019entanglement}%
  \BibitemOpen
  \bibfield  {author} {\bibinfo {author} {\bibfnamefont {K.}~\bibnamefont
  {Selvan}}\ and\ \bibinfo {author} {\bibfnamefont {P.~K.}\ \bibnamefont
  {Panigrahi}},\ }\bibfield  {title} {\bibinfo {title} {Entanglement condition
  for w type multimode states and schemes for experimental realization},\
  }\href@noop {} {\bibfield  {journal} {\bibinfo  {journal} {The European
  Physical Journal D}\ }\textbf {\bibinfo {volume} {73}},\ \bibinfo {pages} {1}
  (\bibinfo {year} {2019})}\BibitemShut {NoStop}%
\bibitem [{\citenamefont {Matthews}\ \emph {et~al.}(2009)\citenamefont
  {Matthews}, \citenamefont {Politi}, \citenamefont {Stefanov},\ and\
  \citenamefont {O'brien}}]{matthews2009manipulation}%
  \BibitemOpen
  \bibfield  {author} {\bibinfo {author} {\bibfnamefont {J.~C.}\ \bibnamefont
  {Matthews}}, \bibinfo {author} {\bibfnamefont {A.}~\bibnamefont {Politi}},
  \bibinfo {author} {\bibfnamefont {A.}~\bibnamefont {Stefanov}},\ and\
  \bibinfo {author} {\bibfnamefont {J.~L.}\ \bibnamefont {O'brien}},\
  }\bibfield  {title} {\bibinfo {title} {Manipulation of multiphoton
  entanglement in waveguide quantum circuits},\ }\href@noop {} {\bibfield
  {journal} {\bibinfo  {journal} {Nature Photonics}\ }\textbf {\bibinfo
  {volume} {3}},\ \bibinfo {pages} {346} (\bibinfo {year} {2009})}\BibitemShut
  {NoStop}%
\bibitem [{\citenamefont {Meher}\ \emph {et~al.}(2017)\citenamefont {Meher},
  \citenamefont {Sivakumar},\ and\ \citenamefont
  {Panigrahi}}]{meher2017duality}%
  \BibitemOpen
  \bibfield  {author} {\bibinfo {author} {\bibfnamefont {N.}~\bibnamefont
  {Meher}}, \bibinfo {author} {\bibfnamefont {S.}~\bibnamefont {Sivakumar}},\
  and\ \bibinfo {author} {\bibfnamefont {P.~K.}\ \bibnamefont {Panigrahi}},\
  }\bibfield  {title} {\bibinfo {title} {Duality and quantum state engineering
  in cavity arrays},\ }\href@noop {} {\bibfield  {journal} {\bibinfo  {journal}
  {Scientific reports}\ }\textbf {\bibinfo {volume} {7}},\ \bibinfo {pages} {1}
  (\bibinfo {year} {2017})}\BibitemShut {NoStop}%
\bibitem [{\citenamefont {Pellizzari}(1997)}]{pellizzari1997quantum}%
  \BibitemOpen
  \bibfield  {author} {\bibinfo {author} {\bibfnamefont {T.}~\bibnamefont
  {Pellizzari}},\ }\bibfield  {title} {\bibinfo {title} {Quantum networking
  with optical fibres},\ }\href@noop {} {\bibfield  {journal} {\bibinfo
  {journal} {Physical Review Letters}\ }\textbf {\bibinfo {volume} {79}},\
  \bibinfo {pages} {5242} (\bibinfo {year} {1997})}\BibitemShut {NoStop}%
\bibitem [{\citenamefont {Van~Enk}\ \emph {et~al.}(1999)\citenamefont
  {Van~Enk}, \citenamefont {Kimble}, \citenamefont {Cirac},\ and\ \citenamefont
  {Zoller}}]{van1999quantum}%
  \BibitemOpen
  \bibfield  {author} {\bibinfo {author} {\bibfnamefont {S.}~\bibnamefont
  {Van~Enk}}, \bibinfo {author} {\bibfnamefont {H.}~\bibnamefont {Kimble}},
  \bibinfo {author} {\bibfnamefont {J.}~\bibnamefont {Cirac}},\ and\ \bibinfo
  {author} {\bibfnamefont {P.}~\bibnamefont {Zoller}},\ }\bibfield  {title}
  {\bibinfo {title} {Quantum communication with dark photons},\ }\href@noop {}
  {\bibfield  {journal} {\bibinfo  {journal} {Physical Review A}\ }\textbf
  {\bibinfo {volume} {59}},\ \bibinfo {pages} {2659} (\bibinfo {year}
  {1999})}\BibitemShut {NoStop}%
\bibitem [{\citenamefont {Lu}\ and\ \citenamefont
  {Chen}(2019)}]{lu2019geometrical}%
  \BibitemOpen
  \bibfield  {author} {\bibinfo {author} {\bibfnamefont {D.-M.}\ \bibnamefont
  {Lu}}\ and\ \bibinfo {author} {\bibfnamefont {L.-H.}\ \bibnamefont {Chen}},\
  }\bibfield  {title} {\bibinfo {title} {Geometrical quantum discord in the
  coupled cavities system with tetrahedral structure},\ }\href@noop {}
  {\bibfield  {journal} {\bibinfo  {journal} {International Journal of
  Theoretical Physics}\ }\textbf {\bibinfo {volume} {58}},\ \bibinfo {pages}
  {605} (\bibinfo {year} {2019})}\BibitemShut {NoStop}%
\bibitem [{\citenamefont {Kyoseva}\ \emph {et~al.}(2012)\citenamefont
  {Kyoseva}, \citenamefont {Beige},\ and\ \citenamefont
  {Kwek}}]{kyoseva2012coherent}%
  \BibitemOpen
  \bibfield  {author} {\bibinfo {author} {\bibfnamefont {E.}~\bibnamefont
  {Kyoseva}}, \bibinfo {author} {\bibfnamefont {A.}~\bibnamefont {Beige}},\
  and\ \bibinfo {author} {\bibfnamefont {L.~C.}\ \bibnamefont {Kwek}},\
  }\bibfield  {title} {\bibinfo {title} {Coherent cavity networks with complete
  connectivity},\ }\href@noop {} {\bibfield  {journal} {\bibinfo  {journal}
  {New Journal of Physics}\ }\textbf {\bibinfo {volume} {14}},\ \bibinfo
  {pages} {023023} (\bibinfo {year} {2012})}\BibitemShut {NoStop}%
\bibitem [{\citenamefont {Cho}\ \emph {et~al.}(2008)\citenamefont {Cho},
  \citenamefont {Angelakis},\ and\ \citenamefont {Bose}}]{cho2008heralded}%
  \BibitemOpen
  \bibfield  {author} {\bibinfo {author} {\bibfnamefont {J.}~\bibnamefont
  {Cho}}, \bibinfo {author} {\bibfnamefont {D.~G.}\ \bibnamefont {Angelakis}},\
  and\ \bibinfo {author} {\bibfnamefont {S.}~\bibnamefont {Bose}},\ }\bibfield
  {title} {\bibinfo {title} {Heralded generation of entanglement with coupled
  cavities},\ }\href@noop {} {\bibfield  {journal} {\bibinfo  {journal}
  {Physical Review A}\ }\textbf {\bibinfo {volume} {78}},\ \bibinfo {pages}
  {022323} (\bibinfo {year} {2008})}\BibitemShut {NoStop}%
\bibitem [{\citenamefont {Bergou}(1999)}]{bergou1999entangled}%
  \BibitemOpen
  \bibfield  {author} {\bibinfo {author} {\bibfnamefont {J.~A.}\ \bibnamefont
  {Bergou}},\ }\bibfield  {title} {\bibinfo {title} {Entangled fields in
  multiple cavities as a testing ground for quantum mechanics},\ }\href@noop {}
  {\bibfield  {journal} {\bibinfo  {journal} {Foundations of physics}\ }\textbf
  {\bibinfo {volume} {29}},\ \bibinfo {pages} {503} (\bibinfo {year}
  {1999})}\BibitemShut {NoStop}%
\bibitem [{\citenamefont {Guo}\ and\ \citenamefont
  {Zhang}(2002)}]{guo2002scheme}%
  \BibitemOpen
  \bibfield  {author} {\bibinfo {author} {\bibfnamefont {G.-C.}\ \bibnamefont
  {Guo}}\ and\ \bibinfo {author} {\bibfnamefont {Y.-S.}\ \bibnamefont
  {Zhang}},\ }\bibfield  {title} {\bibinfo {title} {Scheme for preparation of
  the w state via cavity quantum electrodynamics},\ }\href@noop {} {\bibfield
  {journal} {\bibinfo  {journal} {Physical Review A}\ }\textbf {\bibinfo
  {volume} {65}},\ \bibinfo {pages} {054302} (\bibinfo {year}
  {2002})}\BibitemShut {NoStop}%
\bibitem [{\citenamefont {Yang}\ \emph {et~al.}(2004)\citenamefont {Yang},
  \citenamefont {Yi},\ and\ \citenamefont {Cao}}]{yang2004scheme}%
  \BibitemOpen
  \bibfield  {author} {\bibinfo {author} {\bibfnamefont {M.}~\bibnamefont
  {Yang}}, \bibinfo {author} {\bibfnamefont {Y.-M.}\ \bibnamefont {Yi}},\ and\
  \bibinfo {author} {\bibfnamefont {Z.-L.}\ \bibnamefont {Cao}},\ }\bibfield
  {title} {\bibinfo {title} {Scheme for preparation of w state via cavity
  qed},\ }\href@noop {} {\bibfield  {journal} {\bibinfo  {journal}
  {International Journal of Quantum Information}\ }\textbf {\bibinfo {volume}
  {2}},\ \bibinfo {pages} {231} (\bibinfo {year} {2004})}\BibitemShut {NoStop}%
\bibitem [{\citenamefont {Hillery}\ and\ \citenamefont
  {Mlodinow}(1993)}]{hillery1993interferometers}%
  \BibitemOpen
  \bibfield  {author} {\bibinfo {author} {\bibfnamefont {M.}~\bibnamefont
  {Hillery}}\ and\ \bibinfo {author} {\bibfnamefont {L.}~\bibnamefont
  {Mlodinow}},\ }\bibfield  {title} {\bibinfo {title} {Interferometers and
  minimum-uncertainty states},\ }\href@noop {} {\bibfield  {journal} {\bibinfo
  {journal} {Physical Review A}\ }\textbf {\bibinfo {volume} {48}},\ \bibinfo
  {pages} {1548} (\bibinfo {year} {1993})}\BibitemShut {NoStop}%
\bibitem [{\citenamefont {Lougovski}\ \emph {et~al.}(2009)\citenamefont
  {Lougovski}, \citenamefont {van Enk}, \citenamefont {Choi}, \citenamefont
  {Papp}, \citenamefont {Deng},\ and\ \citenamefont
  {Kimble}}]{lougovski2009verifying}%
  \BibitemOpen
  \bibfield  {author} {\bibinfo {author} {\bibfnamefont {P.}~\bibnamefont
  {Lougovski}}, \bibinfo {author} {\bibfnamefont {S.~J.}\ \bibnamefont {van
  Enk}}, \bibinfo {author} {\bibfnamefont {K.~S.}\ \bibnamefont {Choi}},
  \bibinfo {author} {\bibfnamefont {S.~B.}\ \bibnamefont {Papp}}, \bibinfo
  {author} {\bibfnamefont {H.}~\bibnamefont {Deng}},\ and\ \bibinfo {author}
  {\bibfnamefont {H.}~\bibnamefont {Kimble}},\ }\bibfield  {title} {\bibinfo
  {title} {Verifying multipartite mode entanglement of w states},\ }\href@noop
  {} {\bibfield  {journal} {\bibinfo  {journal} {New Journal of Physics}\
  }\textbf {\bibinfo {volume} {11}},\ \bibinfo {pages} {063029} (\bibinfo
  {year} {2009})}\BibitemShut {NoStop}%
\bibitem [{\citenamefont {Chen}\ \emph {et~al.}(2017)\citenamefont {Chen},
  \citenamefont {Zhou},\ and\ \citenamefont {Shen}}]{chen2017exact}%
  \BibitemOpen
  \bibfield  {author} {\bibinfo {author} {\bibfnamefont {Z.}~\bibnamefont
  {Chen}}, \bibinfo {author} {\bibfnamefont {Y.}~\bibnamefont {Zhou}},\ and\
  \bibinfo {author} {\bibfnamefont {J.-T.}\ \bibnamefont {Shen}},\ }\bibfield
  {title} {\bibinfo {title} {Exact dissipation model for arbitrary photonic
  fock state transport in waveguide qed systems},\ }\href@noop {} {\bibfield
  {journal} {\bibinfo  {journal} {Optics letters}\ }\textbf {\bibinfo {volume}
  {42}},\ \bibinfo {pages} {887} (\bibinfo {year} {2017})}\BibitemShut
  {NoStop}%
\bibitem [{\citenamefont {Chen}\ \emph {et~al.}(2018)\citenamefont {Chen},
  \citenamefont {Zhou},\ and\ \citenamefont {Shen}}]{chen2018entanglement}%
  \BibitemOpen
  \bibfield  {author} {\bibinfo {author} {\bibfnamefont {Z.}~\bibnamefont
  {Chen}}, \bibinfo {author} {\bibfnamefont {Y.}~\bibnamefont {Zhou}},\ and\
  \bibinfo {author} {\bibfnamefont {J.-T.}\ \bibnamefont {Shen}},\ }\bibfield
  {title} {\bibinfo {title} {Entanglement-preserving approach for
  reservoir-induced photonic dissipation in waveguide qed systems},\
  }\href@noop {} {\bibfield  {journal} {\bibinfo  {journal} {Physical Review
  A}\ }\textbf {\bibinfo {volume} {98}},\ \bibinfo {pages} {053830} (\bibinfo
  {year} {2018})}\BibitemShut {NoStop}%
\end{thebibliography}%

\end{document}